\documentclass[11pt]{article}
\usepackage{fullpage}
\usepackage[bookmarks=true,bookmarksnumbered=true,setpagesize=false]{hyperref}
\usepackage{graphicx}
\usepackage{color}
\usepackage{bm}
\usepackage{epsfig}
\usepackage{amsmath,amssymb}
\usepackage{booktabs}
\usepackage{multirow}
\usepackage{physics}

\usepackage{todonotes}

\renewcommand{\L}{\text{L}}
\newcommand{\R}{\text{R}}
\newcommand{\LR}{\text{LR}}
\newcommand{\C}{\text{C}}

\newcommand{\tx}{\text}
\newcommand{\ds}{\displaystyle}
\newcommand{\nn}{\nonumber\\}
\newcommand{\ol}{\overline}
\newcommand{\ov}{\over}

\newcommand{\paren}[1]{\left(#1\right)}
\newcommand{\Paren}[1]{\bigl(#1\bigr)}

\begin{document}
\title{
\begin{flushright}
\ \\*[-80pt] 
\begin{minipage}{0.2\linewidth}
\normalsize
IPMU19-0186\\
OU-HET-1040
\\*[50pt]
\end{minipage}
\end{flushright}
Proton Decay and Axion Dark Matter\\
in $SO(10)$ Grand Unification\\
via Minimal Left-Right Symmetry
\bigskip\\}
\author{
  \centerline{
  	Yuta~Hamada,\footnote{E-mail: \tt hamada@apc.in2p3.fr}
  	\
  	Masahiro~Ibe,\footnote{E-mail: \tt ibe@icrr.u-tokyo.ac.jp}
	\
    Yu~Muramatsu,\footnote{E-mail: \tt yumurasub@gmail.com}}
    \\
    \centerline{
    Kin-ya~Oda,\footnote{E-mail: \tt odakin@phys.sci.osaka-u.ac.jp}
    \
    and
    Norimi~Yokozaki\footnote{E-mail: \tt yokozaki@tuhep.phys.tohoku.ac.jp}
}
  \\[25pt]
  \centerline{
    \begin{minipage}{\linewidth}
      \begin{center}
       $^*${\it \small Universit\'e de Paris, CNRS, Astroparticule et Cosmologie, F-75013 Paris, France}\\
          $^\dagger${\it \small ICRR, University of Tokyo, Kashiwa, Chiba 277-8582, Japan}\\
      $^\dagger${\it \small Kavli IPMU (WPI), UTIAS, The University of Tokyo, Kashiwa, Chiba 277-8583, Japan}\\
       $^\ddagger${\it \small
     Institute~of~Particle~Physics~and~Key~Laboratory~of~Quark~and~Lepton~
Physics~(MOE), Central~China~Normal~University,
~Wuhan,~Hubei~430079,
~People's~Republic~of~China}\\
$^\S${\it \normalsize Department of Physics, Osaka University, Osaka 560-0043, Japan}\\
        $^\P${\it \normalsize Department of Physics, Tohoku University, Sendai, Miyagi 980-8578, Japan
        }\\
      \end{center}
  \end{minipage}}
  \\[50pt]}
\date{}

\maketitle

\begin{abstract}\noindent
We study the proton lifetime in the $SO(10)$ Grand Unified Theory (GUT), which has the left-right (LR) symmetric gauge theory below the GUT scale.
In particular, we focus on the minimal model without the bi-doublet Higgs field in the LR symmetric model,
which predicts the LR-breaking scale at around $10^{10\text{--}12}$\,GeV.
The Wilson coefficients of the proton decay operators turn out to be considerably larger 
than those in the minimal $SU(5)$ GUT model especially when the Standard Model Yukawa interactions are generated
by integrating out extra vector-like multiplets.
As a result, we find that the proton lifetime can be within the reach of the Hyper-Kamiokande experiment even when the
GUT gauge boson mass is in the $10^{16\text{--}17}$\,GeV range.
We also show that the mass of the extra vector-like multiplets can be generated by the 
Peccei-Quinn symmetry breaking in a consistent way with the axion dark matter scenario.
\end{abstract}

\newpage

\section{Introduction}
\label{sec:intro}
The Grand Unified Theory (GUT) \cite{Georgi:1974sy} is one of the most attractive candidates for physics beyond the Standard Model (SM), which provides an explanation of the charge quantization.
In particular, the $SO(10)$ gauge group \cite{SO(10)} is one of the most attractive candidates for the unification group 
as it not only unifies all the gauge interactions in the SM but also unifies 
a generation of the SM fermions into one representation.
Furthermore, it also predicts the existence of the right-handed neutrinos, which naturally explains 
the light active neutrino masses through the seesaw mechanism~\cite{seesaw}.
This feature is a great advantage compared to the $SU(5)$ GUT. 

Another interesting feature of the $SO(10)$ GUT  is that the rank of $SO(10)$ is larger than the SM.
Accordingly, the $SO(10)$ GUT allows various symmetry breaking paths to the SM gauge groups, 
such as the Left-Right (LR) symmetric groups~\cite{LR}.
Among these possibilities, the minimal model based on the $SU(3)_\C\times SU(2)_\L\times SU(2)_\R\times U(1)_{B-L}$ gauge 
group without a bi-doublet of $SU(2)_\L\times SU(2)_\R$
uniquely predicts an intermediate breaking scale of the LR symmetry to be around 
$10^{10\mbox{--}12}$\,GeV~\cite{Chang:1984qr,Siringo:2012bc}; see also \cite{Rajpoot:1980xy,Fukugita:2003en}.
This model also gets renewed attention as it can explain the small Higgs quartic coupling constant
at a high energy scale while solving the strong CP problem simultaneously~\cite{SO10_LR,SO10_LR2}.
In this class of models, all the SM Yukawa interactions are generated by integrating out 
extra vector-like multiplets at around the LR-breaking scale.

In this paper, we discuss the proton lifetime in this scenario with the simplest possibility of the extra matter content.\footnote{
Hall and Harigaya have extensively studied various possibilities of the extra matter contents~\cite{SO10_LR,SO10_LR2}.
In this paper we only study the simplest one among them, which has also been partly discussed in a different context of the axion dark matter scenario~\cite{LR-Unif}.}
As we will see, the preferred GUT scale $\lesssim 10^{17}$\,GeV is lower than expected in Refs.~\cite{Chang:1984qr,Siringo:2012bc} by a factor a few or so,
due to the effects of the extra matter multiplets on the renormalization group running.\footnote{
Recall that the proton decay rate $\Gamma$ is inversely proportional to the fourth power of the GUT gauge-boson mass $M_X$, and hence the slight change of $M_X$ significantly affects $\Gamma$.}
We also find that the Wilson coefficients of the proton decay operators are considerably larger 
than those in the minimal $SU(5)$ GUT model due to the larger gauge coupling below the GUT scale
as well as the $SU(2)_\R$ gauge interaction at the intermediate scale.
As a result, the proton decay rate is enhanced and a parameter region
consistent with the gauge coupling unification in the $10^{16\tx{--}17}$\,GeV range can be tested 
by the Hyper-Kamiokande (Hyper-K) experiment.
We also discuss a possibility to generate the mass of the extra vector-like multiplet by the Peccei-Quinn (PQ) symmetry breaking in a consistent way with the axion dark matter scenario.

The organization of the paper is as follows.
In Sec.~\ref{sec:model}, we summarize the $SO(10)$ model which has the minimal LR-symmetric gauge group at the intermediate stage.
In Sec.~\ref{sec:GCU}, we discuss the gauge coupling unification in the minimal LR symmetric model.
In Sec.~\ref{sec:Proton}, we study the proton lifetime.
In Sec.~\ref{sec:axion}, we discuss the mass generation of the extra vector-like multiplets by 
the PQ symmetry breaking.
We give a summary of our discussion in the final section.

\section{The minimal setup of the $SO(10)$ GUT model}
\label{sec:model}
In this paper, we discuss $SO(10)$ GUT with the following chain of symmetry breaking:
\begin{align}
SO(10)
	\underset{M_{\text{GUT}}}{\longrightarrow}
		G_\LR
			&\equiv SU(3)_\C \times SU(2)_\L \times SU(2)_\R \times U(1)_{B-L} \nn 
	\underset{M_\R}{\longrightarrow}
		G_\tx{SM}
			&\equiv SU(3)_\C \times SU(2)_\L \times U(1)_Y.
\end{align}
To ensure this chain and subsequent SM symmetry breaking, we introduce an $SO(10)$ adjoint Higgs $H_{45}$ and an $SO(10)$ spinor-representation Higgs $H_{16}$.
$H_{16}$ contains the doublet Higgs bosons of  $SU(2)_\R$ and $SU(2)_\L$, respectively.%
\footnote{It should be noted that $H_{45}$ and $H_{16}$ do not contain the bi-doublet Higgs of $SU(2)_\R\times SU(2)_\L$.}
First, the vacuum expectation value (VEV) of $H_{45}$ breaks down the $SO(10)$ symmetry at the GUT scale $M_{\text{GUT}}$. 
Second, the VEV of the $SU(2)_\R$ doublet Higgs breaks down the LR symmetry
at $M_\R$ which we call the LR symmetry breaking scale.
In this setup there is no bi-doublet Higgs.
Below the LR symmetry breaking scale, the $U(1)_Y$ gauge symmetry in the SM is obtained by,
\begin{eqnarray}
Q_Y = \frac{1}{2}Q_{B-L} - T_3^{\R},
\end{eqnarray}
where $T_3^{\R}$ is the third generator of $SU(2)_\R$.
As will be discussed, the typical values of the GUT and the LR symmetry breaking scales are
$M_{\text{GUT}} ={\mathcal O}(10^{16\text{--}17})$\,GeV and $M_R = {\mathcal O}( 10^{10\text{--}12})$\,GeV, respectively.
Throughout this paper, we assume these minimal contents for the Higgs sector,
and assume that only the doublet Higgs bosons of $SU(2)_\R$ and $SU(2)_\L$ remain
massless 
below the GUT scale.

\begin{table}[t]
\centering
\bgroup
\def\arraystretch{2.8}
\begin{tabular}{c||c|c}
 & $G_\LR$ & $G_{\text{SM}}$ \\ \hline \hline
up-type Yukawa & 
$\ds y_u \frac{\paren{{Q_\R}^c H_\R^*}\paren{Q_\L H_\L}}{M_\text{extra}}$ & 
$\ds y_u \frac{\langle H_\R \rangle}{M_\text{extra}} {u_\R}^c\paren{q_\L  h_\L}$ \\ \hline
down-type Yukawa & 
$\ds y_d \frac{\paren{{Q_\R}^c H_\R}\paren{H_\L^* Q_\L}}{M_\text{extra}}$ & 
$\ds y_d \frac{\langle H_\R \rangle}{M_\text{extra}}{d_\R}^c\paren{h_\L^* q_\L}$ \\ \hline
charged Lepton Yukawa & 
$\ds y_d \frac{\paren{{L_\R}^c H_\R}\paren{H_\L^* L_\L}}{M_\text{extra}}$ & 
$\ds y_d \frac{\langle H_\R \rangle}{M_\text{extra}}{e_\R}^c\paren{h_\L^* l_\L}$ \\ \hline
neutrino Dirac Yukawa & 
$\ds y_u \frac{\paren{{L_\R}^c H_\R^*}\paren{L_\L H_\L}}{M_\text{extra}}$ & 
$\ds y_u \frac{\langle H_\R \rangle}{M_\text{extra}} {\nu_\R}^c\paren{l_\L  h_\L}$ \\ \hline
neutrino Majorana mass & 
$\ds y_u \frac{\paren{{L_\R}^c H_\R^*}\paren{{L_\R}^c H_\R^*}}{M_\text{extra}}$ & 
$\ds y_u \frac{\langle H_\R \rangle^2}{M_\text{extra}} {\nu_\R}^c {\nu_\R}^c$ \\ \hline
\end{tabular}
\egroup
\caption{\sl \small The Yukawa interactions which come from the higher dimensional operators in Eq.\,(\ref{Eq:Yukawa_higher}).}
\label{Tab:Yukawa}
\end{table}

In the minimal $SO(10)$ GUT model, each generation of the quarks and the leptons of the SM forms 
an $SO(10)$-spinor $F_{16}$, which is decomposed into the $G_\LR$ and $G_\tx{SM}$ representations as
\begin{align}
F_{16} 
	&\underset{M_{\text{GUT}}}{\longrightarrow}
		Q_\L(3,2,1)_{1\ov3}+{Q_\R}^c(\ol 3,1,2)_{-{1\ov3}}+
		L_\L(1,2,1)_{-1}+{L_\R}^c(1,1,2)_1\nn
	&\underset{M_\R}{\longrightarrow}
		q_\L(3,2)_{1\ov6}
		+\paren{{d_\R}^c(\ol 3,1)_{1\ov3}+{u_\R}^c(\ol 3,1)_{-{2\ov3}}}
		+l_\L(1,2)_{-{1\ov2}}
		+\Paren{{e_\R}^c(1,1)_1+{\nu_\R}^c(1,1)_0} ,
\end{align}
where the subscript is the charges of $B-L$ and $Y$, respectively.
To embed $U(1)_{B-L}$ into $SO(10)$, we renormalize the charges so that the $U(1)$ gauge couplings are given by%
\footnote{With these normalizations, the Dynkin index of the $U(1)$ gauge group of $F_{16}$ becomes 
$2$.} 
\begin{align}
\alpha_Y
	&=	{3\ov5}\alpha_1\quad\tx{(below $M_\R$)},&
\alpha_{B-L}
	&=	{3\ov8}\alpha_1\quad\tx{(above $M_\R$)}.
\end{align}

In the LR symmetric model with only $SU(2)_{L,R}$ doublet Higgs bosons, 
the Yukawa interactions in the SM are given by the higher dimensional operators
in Table~\ref{Tab:Yukawa}.
In the $SO(10)$ notation, they correspond to 
\begin{equation}
\mathcal{L}_Y=\frac{y_{u\,ij}}{\Lambda} \paren{F_{16\,i}H^*_{16}} \paren{F_{16\,j}  H^*_{16} }
+ \frac{y_{d\,ij}}{\Lambda} \paren{F_{16\,i}H_{16}}\paren{F_{16\,j}  H_{16}} + \text{h.c},
\label{Eq:Yukawa_higher}
\end{equation}
where $i,j=1,2,3$ is the flavor index. $\Lambda$ is the cutoff scale.
Hereafter, we suppress the gauge and  the flavor indices unless otherwise stated.
After the LR symmetry breaking, these operators contribute to the Yukawa interactions:
 $y_u$ contributes to the up-type and neutrino ones, while $y_d$ to the down-type and charged-lepton ones.
In Table~\ref{Tab:Yukawa}, the second and the third columns represents 
the Yukawa interactions from the higher dimensional operators in Eq.\,(\ref{Eq:Yukawa_higher}) 
in the representations of $G_\LR$ and $G_{\text{SM}}$, respectively.

Obviously, these contributions are too small to realize the observed masses of the heavy flavor fermions in the SM for 
$\Lambda = M_{\text{GUT}}$, for example.
In fact, since the LR symmetry breaking scale $M_\R$ is around $10^{10}$--$10^{11}$\,GeV,
while $M_{\text{GUT}} =10^{16}$--$10^{17}$\,GeV,
the coefficient of these operators are $ \sim M_\R / \Lambda = 10^{-7}$--$10^{-5}$, and hence we cannot realize the Yukawa couplings for the second and third generations.
To reproduce the observed quark and lepton masses, we need to introduce extra vector-like multiplets 
with masses of ${M_\R}$ so that the terms in Eq.\,\eqref{Eq:Yukawa_higher} are generated by 
integrating out those extra multiplets.

In this paper, we assume that all the SM Yukawa interactions are generated by integrating out extra vector-like multiplets.
In this case, the minimal extra vector-like fermions consist of three flavors of the fundamental representation of $SO(10)$, $E_{10}$,
and three flavors of the adjoint representation of $SO(10)$, $E_{45}$.%
\footnote{See \cite{SO10_LR,SO10_LR2} for other possibilities.}.
When the Yukawa interactions of the first generation are provided by the $M_{\text{GUT}}$ suppressed
operators, two flavors of $E_{10}$ and $E_{45}$ are enough to reproduce the SM Yukawa interactions.
As discussed in Sec.~\ref{sec:axion}, however, the three flavor model is advantageous as the masses 
of the extra vector-like fermions can be interrelated to the PQ symmetry breaking.
In what follows, we denote the number of extra particle flavors by $N_E$.

With these extra matter multiplets, the origin of the  Yukawa interactions in Eq.\,\eqref{Eq:Yukawa_higher} are obtained 
from the renormalizable interactions,
\begin{equation}
\mathcal{L}_{\text{extra}} = y'_d F_{16} E_{10} H_{16} + y'_u F_{16} E_{45} H^*_{16} + M_{\text{extra}} E_{10} E_{10} + M_{\text{extra}} E_{45} E_{45} + \tx{h.c,}
\label{Eq:extra_cont}
\end{equation}
where $M_{\text{extra}}$ is the extra particle mass. 
We assume that the mass parameters for $E_{10}$ and $E_{45}$ are the same for simplicity. 
$E_{10}$ and $E_{45}$ are decomposed into the $G_\LR$ representations as 
\begin{align}
E_{10} 
	&\underset{M_{\text{GUT}}}{\longrightarrow}
		D^{(10)}(3,1,1)_{-\frac{2}{3}} + \overline{D}^{(10)}(\overline{3},1,1)_{\frac{2}{3}} + L_\LR^{(10)}(1,2,2)_0,\\
E_{45} 
	&\underset{M_{\text{GUT}}}{\longrightarrow}
		W_{L}^{(45)}(1,3,1)_0 + W_\R^{(45)}(1,1,3)_0 + G^{(45)}(8,1,1)_0 + N^{(45)}(1,1,1)_0 \nonumber \\
	&\phantom{\underset{M_{\text{GUT}}}{\longrightarrow}}
		+U^{(45)}(3,1,1)_{\frac{4}{3}} + \overline{U}^{(45)}(\overline{3},1,1)_{-\frac{4}{3}} + Q_\LR^{(45)}(3,2,2)_{-\frac{2}{3}} + \overline{Q}_\LR^{(45)}(\overline{3},2,2)_{\frac{2}{3}}.
\end{align}
Here and hereafter, the overline on the extra fields denotes a charge conjugation rather than a Dirac adjoint.
By using the $G_\LR$ representations, the Eq.\,(\ref{Eq:extra_cont}) is decomposed as
\begin{align}
\mathcal{L}_{\text{extra}}
	&\supset
		y'_d Q_\L H_\L^* \overline{D}^{(10)} + y'_d {L_\R}^c H_\L^* L_\LR^{(10)}\nonumber \\
	&\quad
		+y'_d {Q_\R}^c H_\R D^{(10)} + y'_d L_\L H_\R L_\LR^{(10)} \nonumber \\
	&\quad
		+y'_u Q_\L H_\L \overline{U}^{(45)} + y'_u L_\L H_\L N^{(45)} \nonumber \\
	&\quad
		+y'_u {Q_\R}^c H_\R^* U^{(45)} + y'_u {L_\R}^c H_\R^* N^{(45)}.
\end{align}
When we integrate out the extra particles, these contributions become the higher dimensional operators which are summarized in Table~\ref{Tab:Yukawa}.
The resultant Yukawa coupling constants in the SM are proportional to $M_\R / M_{\text{extra}}$, 
and hence, 
the top Yukawa coupling requires the extra particle masses should be around LR symmetry breaking scale.

Several comments of the above minimal setup is summarized as follow; see also \cite{SO10_LR,SO10_LR2}.
\begin{itemize}
\item The large difference between the top mass and the other third generation one is the most serious problem in realization of the observed fermion masses in generic $SO(10)$ GUT models.
This is not a matter in our model because there are two origins of the Yukawa interactions.
\item Small difference between down-type quark masses and charged-lepton masses is introduced by higher dimensional operators that come from $SO(10)$ breaking effects~\cite{Ellis:1979fg}.
\end{itemize}
\begin{itemize}
\item The right-handed neutrino masses are around LR-breaking scale $M_\R$.
As we assume that the LR-breaking scale is around $10^{10}$--$10^{12}$\,GeV, 
the masses of the active neutrinos generated by the seesaw mechanism~\cite{seesaw}
tend to be much heavier than the observed ones. 
This is because the Dirac neutrino Yukawa coupling for third-generation is $O(1)$ 
since it is unified with the top Yukawa coupling.
\item 
When we assume that large mixings in the MNS matrix are realized, the CKM matrix also should be  a 
large mixing matrix because of the unification. However, this does not satisfy experimental results.
\end{itemize}
We can solve the above problems by cancellation between the contributions from the operators 
in Eq.\,(\ref{Eq:Yukawa_higher}) with some other higher-dimensional operators which include 
the GUT breaking effects.
The latter operators are suppressed by a factor of $\langle H_{45} \rangle / \Lambda$. 
However, in this model, the suppression factor is not so small even if $\Lambda$ is around the Planck scale, 
$M_{\mathrm{Pl}}\sim 2.4\times 10^{18}
$\,GeV, 
as $\langle H_{45} \rangle$ can be as large as around $10^{17}$\,GeV. 
By the cancellation, the small neutrino Yukawa coupling can be achieved even for the $O(1)$ top Yukawa coupling, 
and hence the active neutrino masses satisfy the experimental results.
The mixing matrices of the quarks and the neutrinos can also be consistent with each other 
by cancellation.

\section{Gauge Coupling Unification}
\label{sec:GCU}
In the previous section, we introduce extra fermions to achieve the Yukawa interactions of the SM.
In this section, we consider the renormalization group (RG) flow of the gauge couplings 
including the contributions of those extra matter multiplets as well as the $SU(2)_\R$ doublet Higgs boson.
We assume for simplicity that the masses of the extra fermions and $SU(2)_\R$ doublet Higgs are $M_\R$.
As the extra fermions makes the gauge coupling constants become rather strong at around the GUT scale, 
it is important to take into account the two-loop contributions of the gauge coupling constants 
to the RG flow; see e.g.\ Refs.~\cite{GCU}.
The extra Yukawa interactions to the two-loop RGE may slightly affect the 
precision of the unification and the 
GUT gauge boson mass.
Since those effects depend on the 
detailed mass spectrum of the extra fermions, 
we neglect those contributions in this paper.

The $\beta$ function of the gauge coupling~$g_a$ is given by
\begin{align}
\beta_{g_a}&=\frac{1}{16 \pi^2} a_a g_a^3 +  \frac{1}{(16 \pi^2)^2} b_{ab} g_a^3 g_b^2,
\end{align}
where $a,b$ take values $1,2,3$ which refer to $U(1)_{\text{Y}}$, $SU(2)_\L$, and $SU(3)_\C$ below $M_\R$ and take values 1, $2\L$, $2\R$, and 3 which refer to  $U(1)_{B-L}$, $SU(2)_\L$, $SU(2)_\R$, and $SU(3)_\C$ above $M_\R$, respectively:
\begin{itemize}
\item Above $M_\R$, the coefficients of the gauge coupling beta functions are 
\begin{align}
a_a &= \paren{a_0}_a + N_E \paren{a_{10}}_a + N_E \paren{a_{45}}_a,&
b_{ab} &= \paren{b_0}_{ab} + N_E \paren{b_{10}}_{ab} + N_E \paren{b_{45}}_{ab},
\label{eq:beta_coe1}
\end{align}
where each of $a_0$ and $b_0$ contains contributions from the SM particles and the $SU(2)_\R$ doublet Higgs; $a_{10}$ and $b_{10}$ from $E_{10}$;  $a_{45}$ and $b_{45}$ from $E_{45}$; and $N_E$ is the number of extra particle pairs.
These coefficients above $M_R$ are given by\footnote{
There were minor errors 
in the two-loop coefficients in the first arXiv versions of our work and of Ref.~\cite{SO10_LR2}.
The coefficients of the current version have been cross-checked by the authors of Ref.~\cite{SO10_LR2}.
}
\bgroup
\def\arraystretch{1.5}
\begin{align}
\paren{a_0}_a &= \begin{pmatrix}
\frac{9}{2} \\
-\frac{19}{6} \\
-\frac{19}{6} \\
-7
\end{pmatrix},&
\paren{a_{10}}_a &= \begin{pmatrix}
\frac{2}{3} \\
\frac{2}{3} \\
\frac{2}{3} \\
\frac{2}{3}
\end{pmatrix},&
\paren{a_{45}}_a &= \begin{pmatrix}
\frac{16}{3} \\
\frac{16}{3} \\
\frac{16}{3} \\
\frac{16}{3}
\end{pmatrix},\\
\paren{b_0}_{ab} &= \begin{pmatrix}
\frac{23}{4} & \frac{27}{4} & \frac{27}{4} & 4 \\
\frac{9}{4} & \frac{35}{6} & 0 & 12 \\
\frac{9}{4} & 0 & \frac{35}{6} & 12 \\
\frac{1}{2} & \frac{9}{2} & \frac{9}{2} & -26 
\end{pmatrix},&
\paren{b_{10}}_{ab} &= \begin{pmatrix}
\frac{1}{3} & 0 & 0 & \frac{8}{3} \\
0 & \frac{49}{6} & \frac{3}{2} & 0 \\
0 & \frac{3}{2} & \frac{49}{6} & 0 \\
\frac{1}{3} & 0 & 0 & \frac{38}{3}
\end{pmatrix},&
\paren{b_{45}}_{ab} &= \begin{pmatrix}
\frac{20}{3} & 6 & 6 & \frac{64}{3} \\
2 & \frac{211}{3} & 9 & 16 \\
2 & 9 & \frac{211}{3} & 16 \\
\frac{8}{3} & 6 & 6 & \frac{334}{3}
\end{pmatrix}.
\end{align}
We set the $SU(2)_\L$ and $SU(2)_\R$ gauge couplings equal: $g_{2\L}=g_{2\R}\equiv g_2$.
\item Below $M_\R$, on the other hand, they are given by
\begin{align}
a_a &= \begin{pmatrix}
\frac{41}{10} \\
-\frac{19}{6} \\
-7
\end{pmatrix},&
b_{ab} &= \begin{pmatrix}
\frac{199}{50} & \frac{27}{10} & \frac{44}{5} \\
\frac{9}{10} & \frac{35}{6} & 12 \\
\frac{11}{10} & \frac{9}{2} & -26 
\end{pmatrix},
\label{eq:beta_coe2}
\end{align}
\egroup
which come only from the SM particle contribution.\footnote{
Some of our two-loop coefficients differ from those in Ref.~\cite{SO10_LR2} but consistent with those in Ref.~\cite{GCU}.
}
\end{itemize}

To calculate the RG flow for the gauge couplings, we consider the one-loop matching condition at the renormalization scale,
\begin{equation}
\left.\frac{1}{\alpha_1(M_R)}\right|_\tx{Below $M_\R$}
=\left.\frac{3}{5} \frac{1}{\alpha_{2\R}(M_\R)} + \frac{2}{5} \frac{1}{\alpha_1(M_\R)}\right|_\tx{Above $M_\R$}
-\frac{1}{2\pi}\frac{1}{10}.
\end{equation}
Recall that the value of gauge coupling for $SU(2)_\R$ group is the same as that for $SU(2)_\L$ group above $M_\R$: $\alpha_{2\R} = \alpha_{2\L} \equiv \alpha_2$.
As we are taking the $\overline{\text{MS}}$ renormalization scheme, there 
is a mass independent threshold correction in the right-hand side~\cite{Weinberg:1980wa}.%
\footnote{For the $\overline{\text{DR}}$ renormalization scheme, the mass independent threshold correction is absent.}
In the following, we assume that the massive gauge boson of $SU(2)_\R\times U(1)_{B-L}$
and the extra matter multiplets $E_{10,45}$ have the same mass of $M_\R$ for simplicity.
The contributions of the extra matter do not affect the quality of the unification significantly 
as long as they have $SO(10)$ consistent masses.%
\footnote{The mass scales of the extra vector-like matter affect the size of the GUT gauge coupling constant.
The mass splitting within the extra $SO(10)$ multiplets also affect the precision of the gauge 
coupling unification at the GUT scale.}

\begin{figure}[tp]
\centering
\includegraphics[width=5.4cm,pagebox=cropbox,clip]{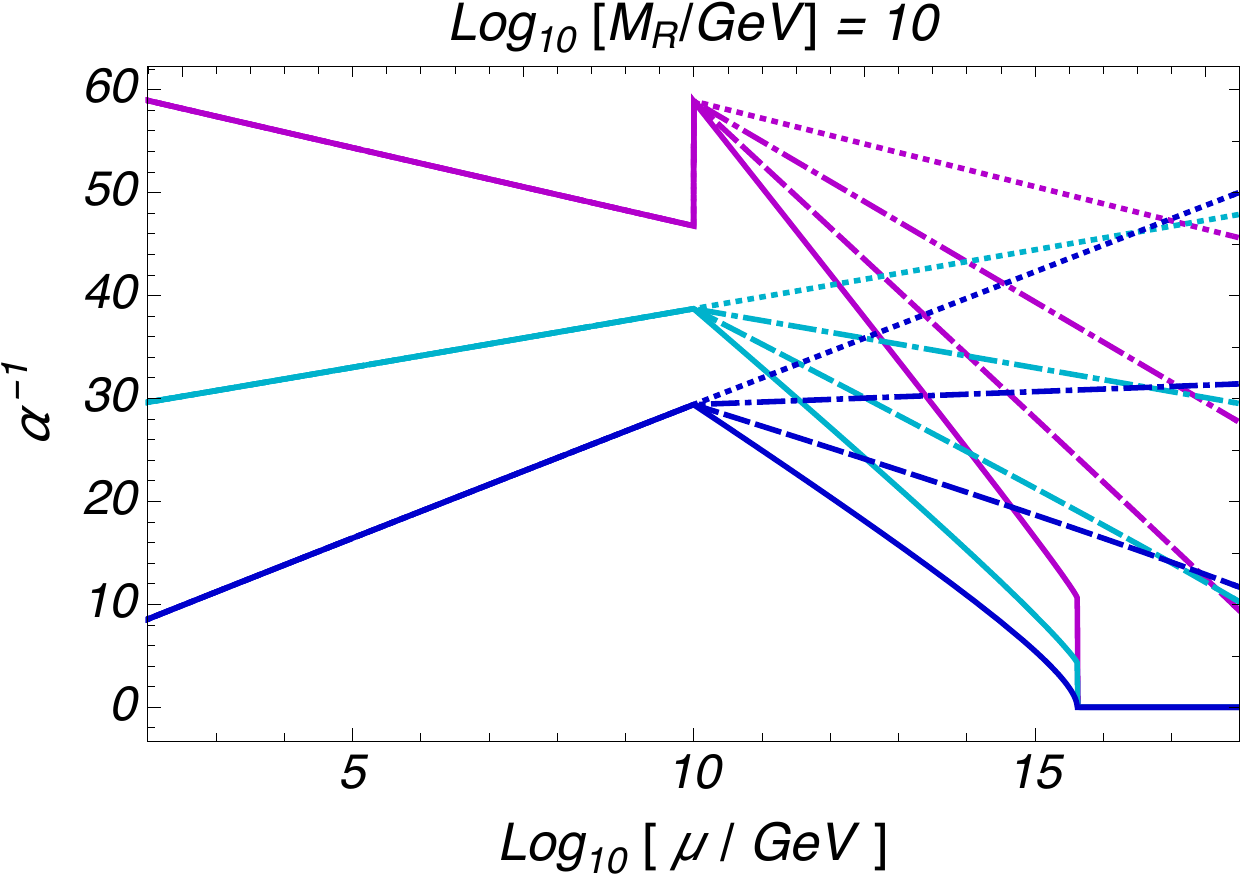}
\includegraphics[width=5.4cm,pagebox=cropbox,clip]{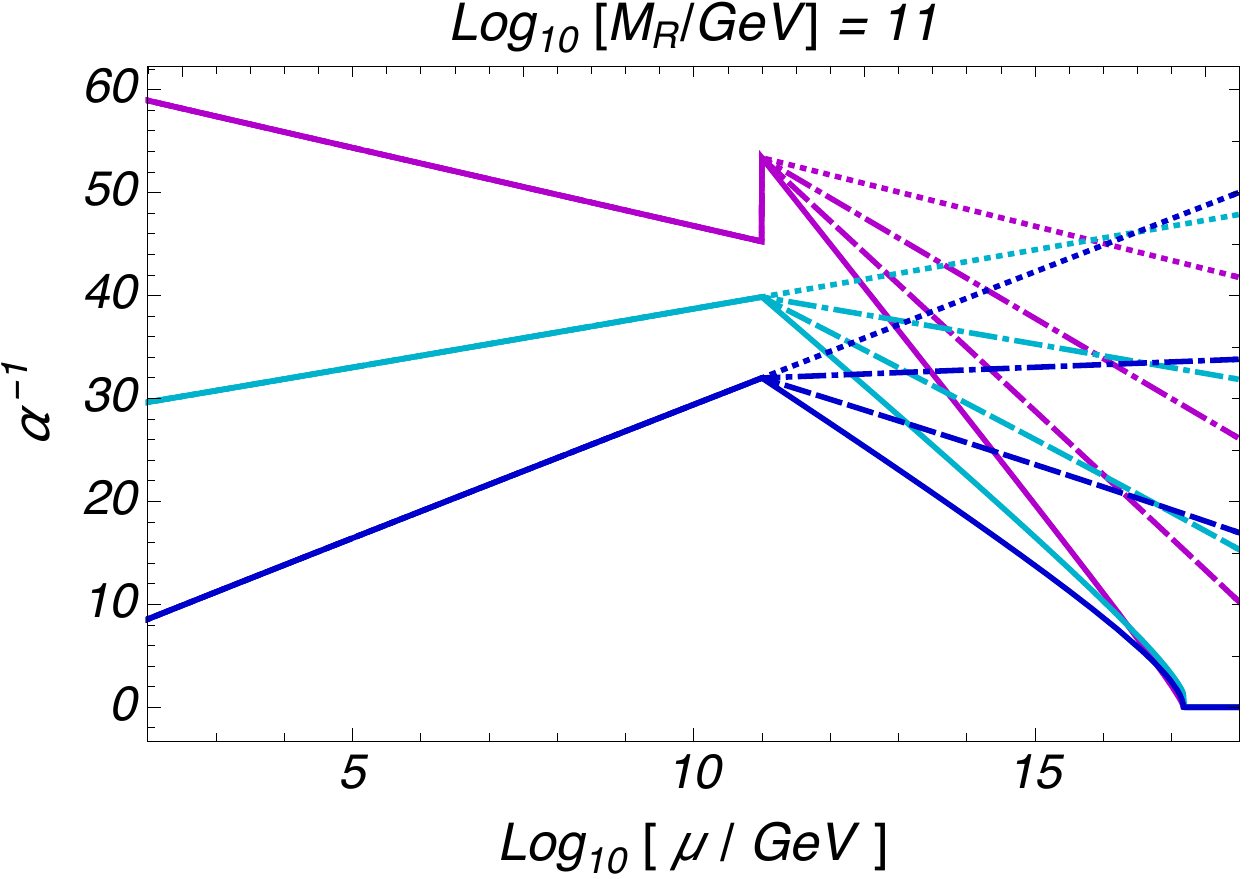}
\includegraphics[width=5.4cm,pagebox=cropbox,clip]{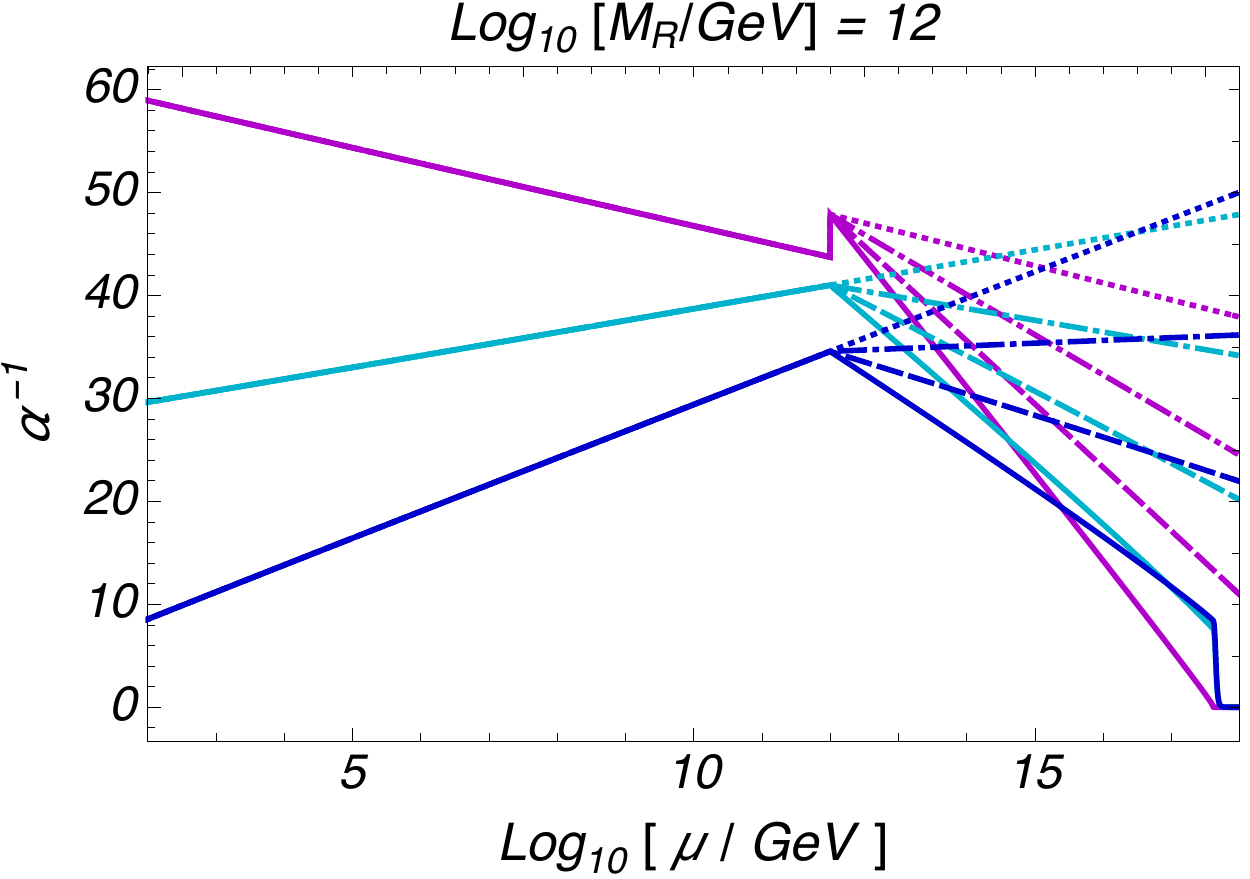}
\caption{\sl \small The RG flow of the gauge couplings for $M_R =10^{10}$\,GeV, $M_R = 10^{11}$\,GeV
and $M_\R = 10^{12}$\,GeV form left to right.
Below the LR symmetry breaking scale $M_\R$, the purple, light-blue, and blue lines refer to gauge couplings 
of  $U(1)_Y$, $SU(2)_\L$, and $SU(3)_\C$ gauge groups, respectively.
Above $M_\R$, the purple line refers to the gauge coupling of the $U(1)_{B-L}$ group.
$N_E =3$ (solid), $N_E=2$ (dashed), $N_E = 2$ (dashed-dotted), and $N_{\text{E} } = 0$ (dotted) 
pairs of the extra fermion are introduced.
}
\label{fig:GCU}
\end{figure}

In Figure \ref{fig:GCU}, the RG flow of the gauge couplings is shown.
The input values for the RG flow are taken to be the central values of the experimental measurements in~\cite{Patrignani:2016xqp}:
\begin{center}
\begin{tabular}{cccc}
$\alpha (M_W)$ & $\alpha_3 (M_Z)$ & $\sin^2 \theta_W(M_Z)$ & $M_Z$ [GeV] \\ \hline
1/128 & 0.1181 & 0.23122 & 91.1876
\end{tabular}
\end{center}
Below the LR symmetry breaking scale $M_\R$, the purple, light-blue, and the blue lines refer to 
the gauge couplings for the $U(1)_\text{Y}$,  the $SU(2)_\L$, and  the $SU(3)_\C$ groups, respectively.
Above $M_\R$, the purple line refers to the gauge coupling for the $U(1)_{B-L}$ group.
$N_E= 3$ (solid), $N_E= 2$ (dashed),  $N_E= 1$ (dashed-dotted), and $N_E= 0$ (dotted) of the 
extra fermions are introduced.
From the left to right, we take the LR symmetry breaking scale, $M_\R = 10^{10}$\,GeV, $M_\R = 10^{11}$\,GeV, 
and $M_\R = 10^{12}$\,GeV, respectively.

The figure shows that the gauge couplings of the LR symmetric model become close 
with each other at around $10^{17}$--$10^{18}$\,GeV for $M_\R = 10^{10}$\,GeV 
for $N_E\le 2$.
The three pairs of the extra multiplets at $M_\R = 10^{10}$\,GeV,
on the other hand, leads to the Landau pole before unification.
The gauge couplings for $M_\R = 10^{11}$\,GeV, on the other hand, meet well together before they hit the  Landau pole.
There, we see that the two-loop contributions are not negligible with which the RG flow becomes non-linear.
The results for $M_\R=10^{12}$\,GeV also show that the gauge couplings become close with each other 
moderately at around $M_\R = 10^{15}$\,GeV.

To quantify the quality of the unification, let us consider the matching conditions between
the gauge coupling constants in the LR symmetric model and the $SO(10)$ gauge coupling, $\alpha_G = g_G^2/4\pi$:
\begin{align}
&\frac{1}{\alpha_{1}(\mu,M_R)} = \frac{1}{\alpha_{G}(\Lambda)} 
- \frac{1}{2\pi} 
\left(a_1 \log\frac{\mu}{\Lambda} - {14} \log\frac{M_X}{\Lambda} 
- {14} \log\frac{M_{X'}}{\Lambda} 
\right)
-\frac{1}{2\pi}\frac{4}{3}
+\frac{1}{2\pi} \Delta_{1},\\
&\frac{1}{\alpha_2(\mu,M_R)} = \frac{1}{\alpha_{G}(\Lambda)} 
- \frac{1}{2\pi}
\left(a_2 \log\frac{\mu}{\Lambda} 
-{21} \log\frac{M_X}{\Lambda} 
\right)
- \frac{1}{2\pi}
 + \frac{1}{2\pi} \Delta_{2} ,\\
&\frac{1}{\alpha_3(\mu,M_R)} = \frac{1}{\alpha_{G}(\Lambda)} 
-\frac{1}{2\pi}
\left(a_3 \log\frac{\mu}{\Lambda} 
- {14} \log\frac{M_X}{\Lambda} 
-  \frac{7}{2}\log\frac{M_{X'}}{\Lambda} 
\right)
-\frac{1}{2\pi}\frac{5}{6}
+\frac{1}{2\pi} \Delta_{3}.
\end{align}
The parameters $\mu$ and $\Lambda$ are the renormalization scale and the cutoff scale
at around the GUT scale.
The mass parameter $M_X$ and $M_{X'}$ denotes the mass of the gauge boson in the $(3,2,2)_{-2/3}$ and $(3,1,1)_{-4/3}$
representations, respectively.
For the symmetry breaking path $SO(10)\to SU(3)\times SU(2)_\L\times SU(2)_{\R}\times U(1)_{B-L}$ 
by the VEV of the Higgs boson in the 45 representation, it is predicted that  $M_{X'} = 2 M_{X}$.
The mass independent threshold corrections are due to the $\overline{MS}$ renormalization scheme, which 
are absent in the $\overline{DR}$ renormalization scheme.
The parameters ${\Delta}_{1,2,3}$ represent the threshold corrections from some particles at 
the GUT scale other than the GUT gauge bosons, although we do not specify them in this paper.%
\footnote{The parameters $\Delta_{1,2,3}$
also get contributions from higher dimensional operators $\langle H_{45}\rangle/M_{\mathrm{Pl}}$,
although we assume that $\langle H_{45}\rangle/M_{\mathrm{Pl}} \ll \order{1}$. }
See \cite{SO10_LR,SO10_LR2} for various contributions of the GUT multiplets to $\Delta_{1,2,3}$.

\begin{figure}[t]
\centering
  \begin{minipage}{.24\linewidth}
 \includegraphics[width=\linewidth]{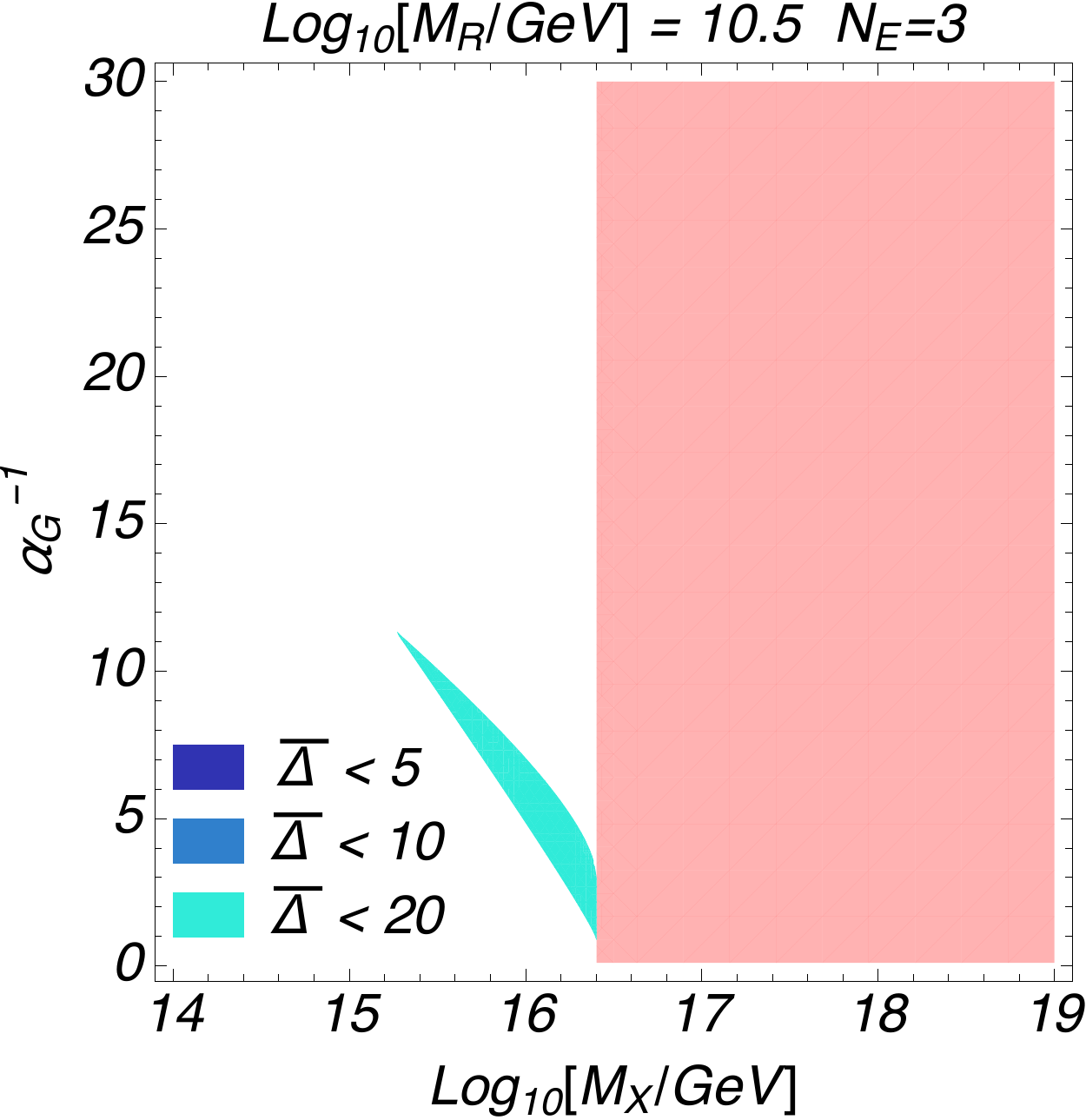}
  \end{minipage}
 \begin{minipage}{.24\linewidth}
 \includegraphics[width=\linewidth]{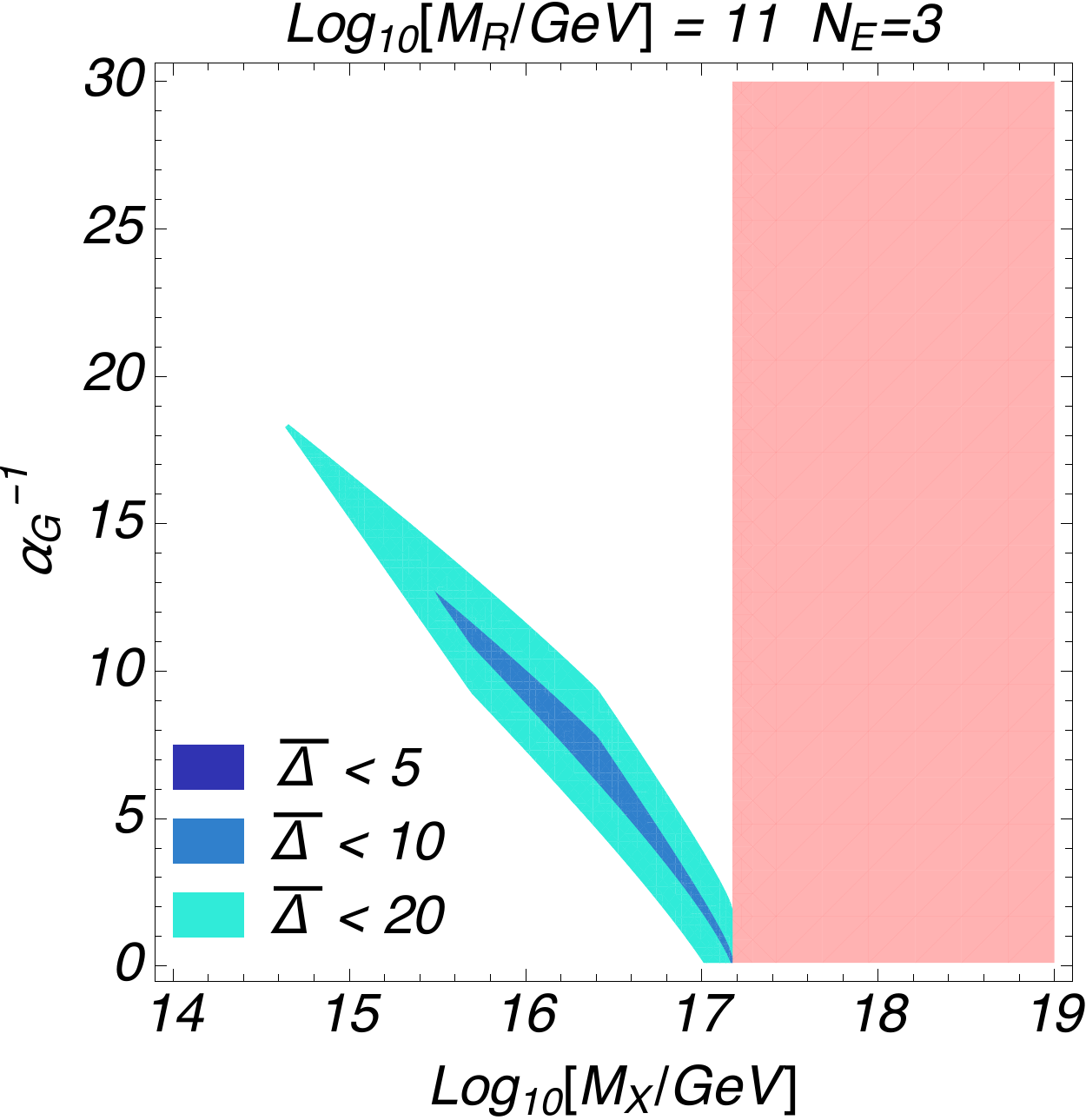}
  \end{minipage}
  \vspace{.5cm}
    \begin{minipage}{.24\linewidth}
 \includegraphics[width=\linewidth]{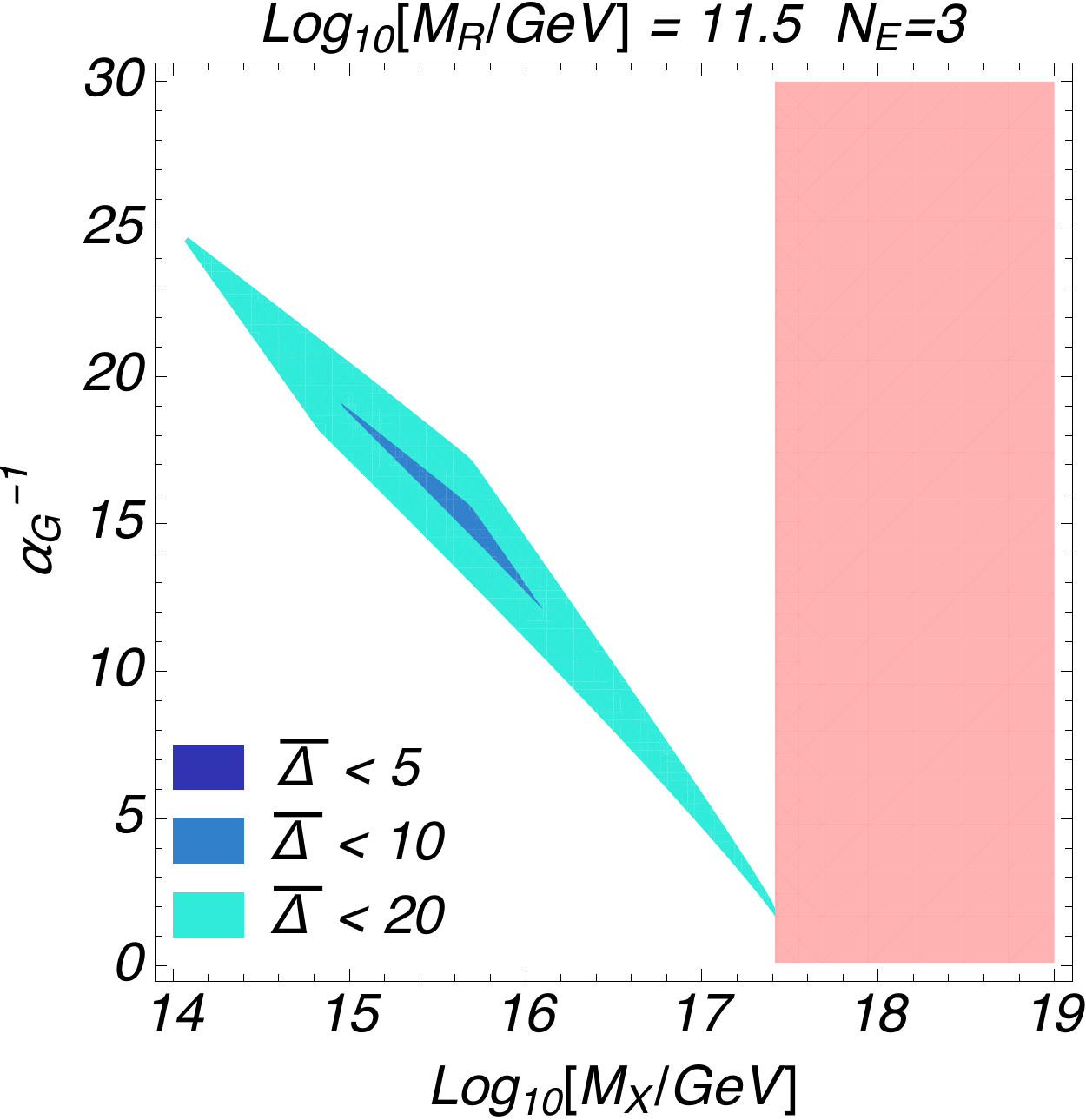}
  \end{minipage}
    \begin{minipage}{.24\linewidth}
 \includegraphics[width=\linewidth]{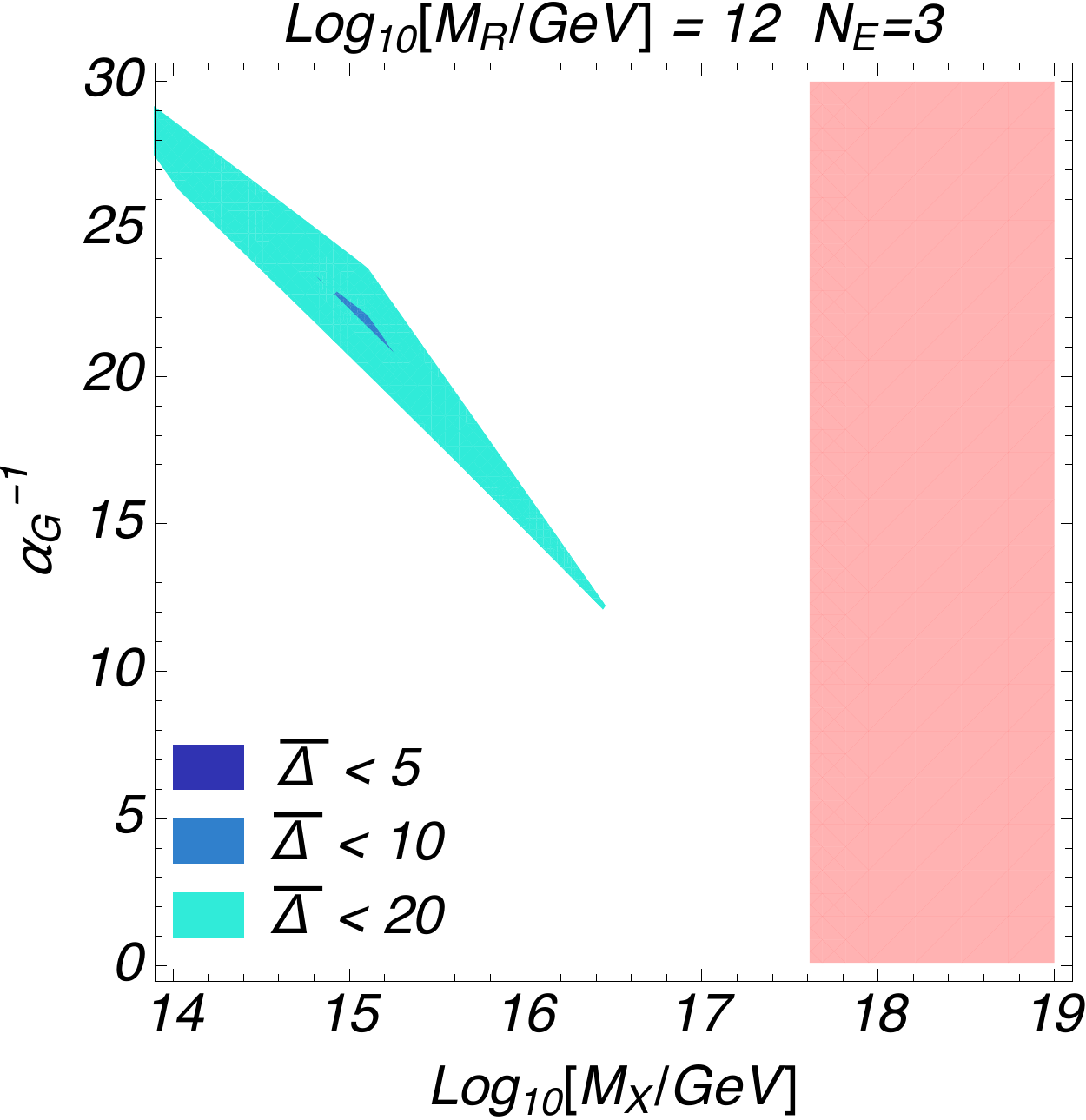}
   \end{minipage}
     \begin{minipage}{.24\linewidth}
 \includegraphics[width=\linewidth]{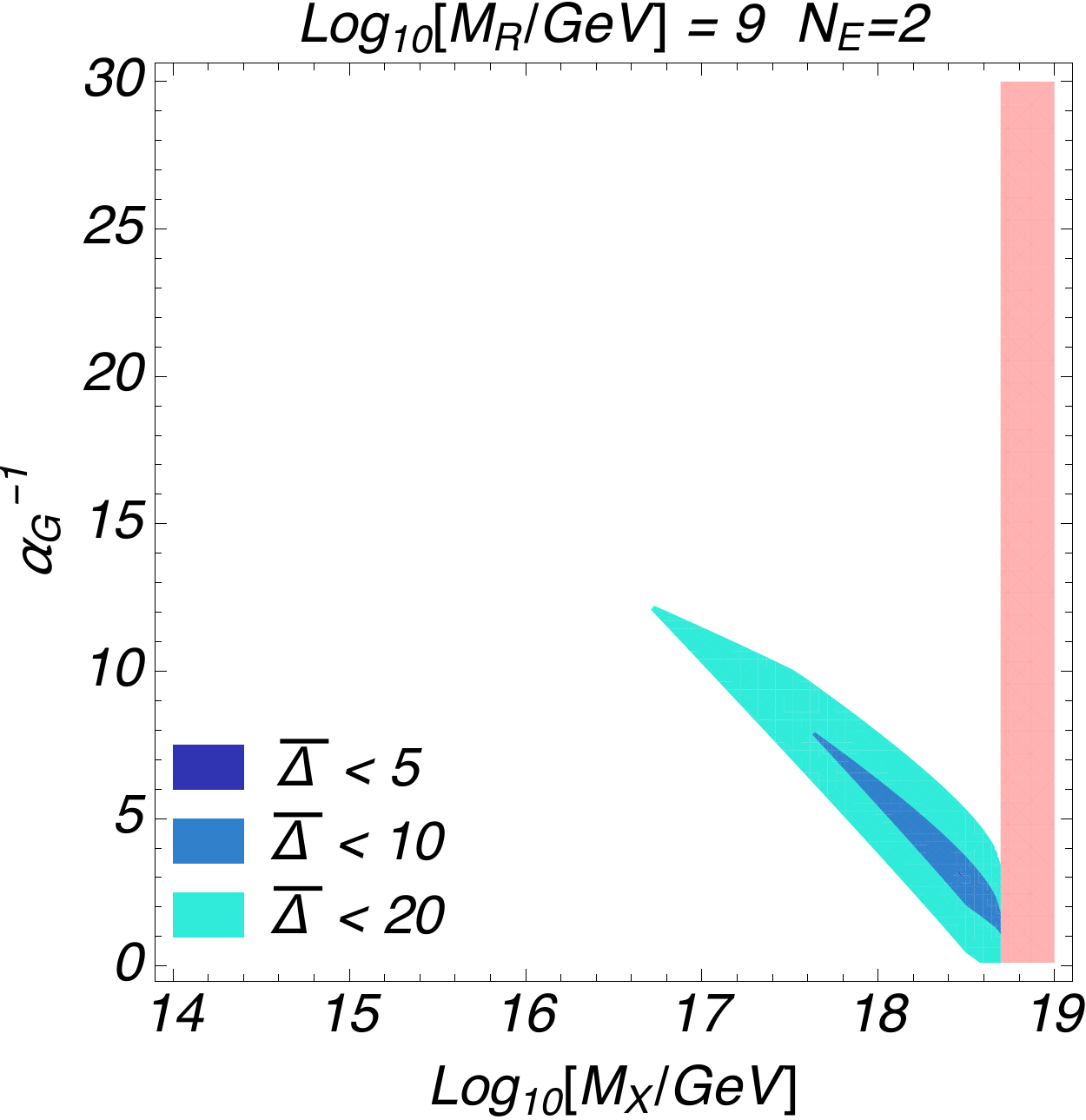}
  \end{minipage}
 \begin{minipage}{.24\linewidth}
 \includegraphics[width=\linewidth]{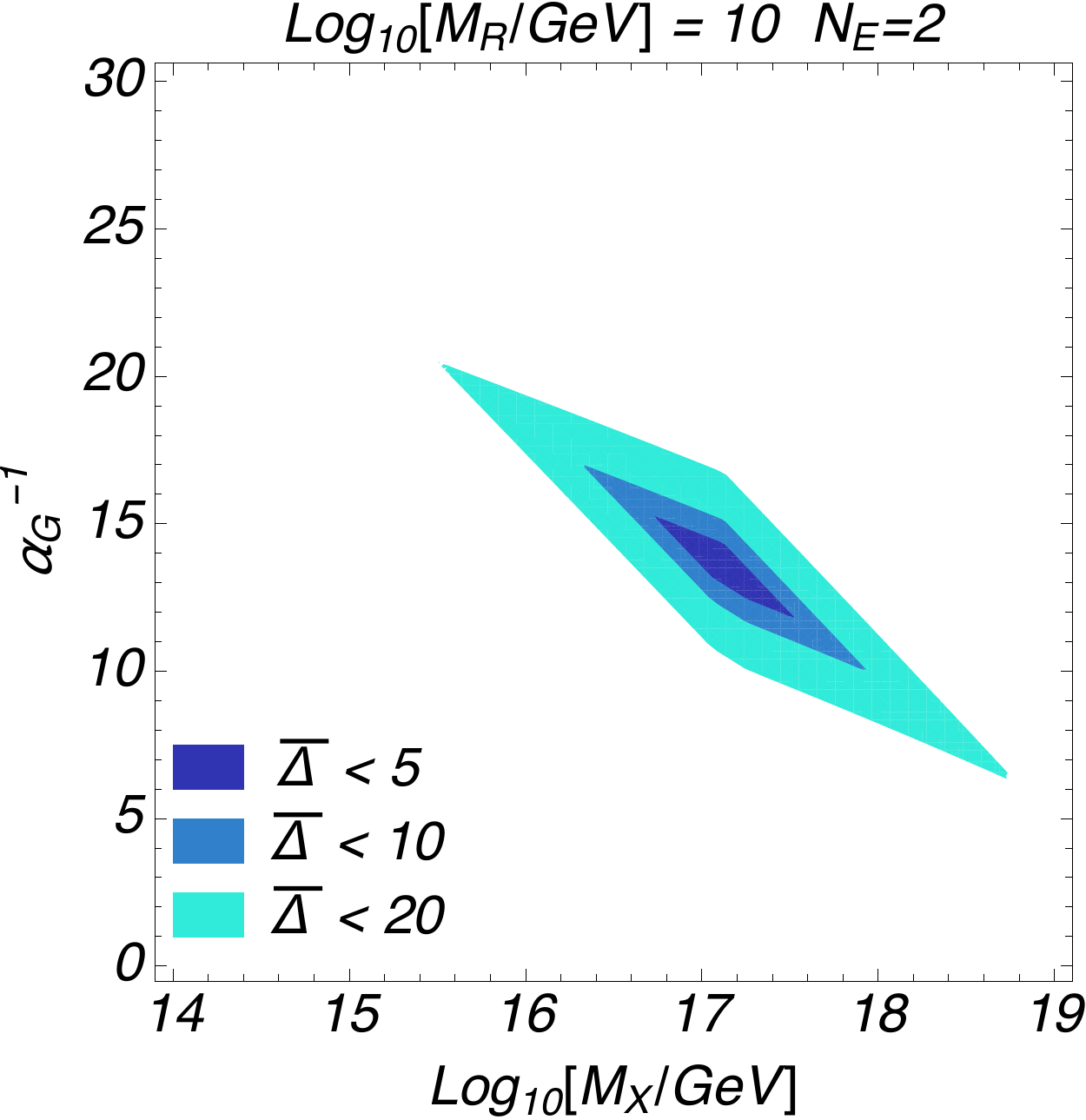}
  \end{minipage}
  \vspace{.5cm}
    \begin{minipage}{.24\linewidth}
 \includegraphics[width=\linewidth]{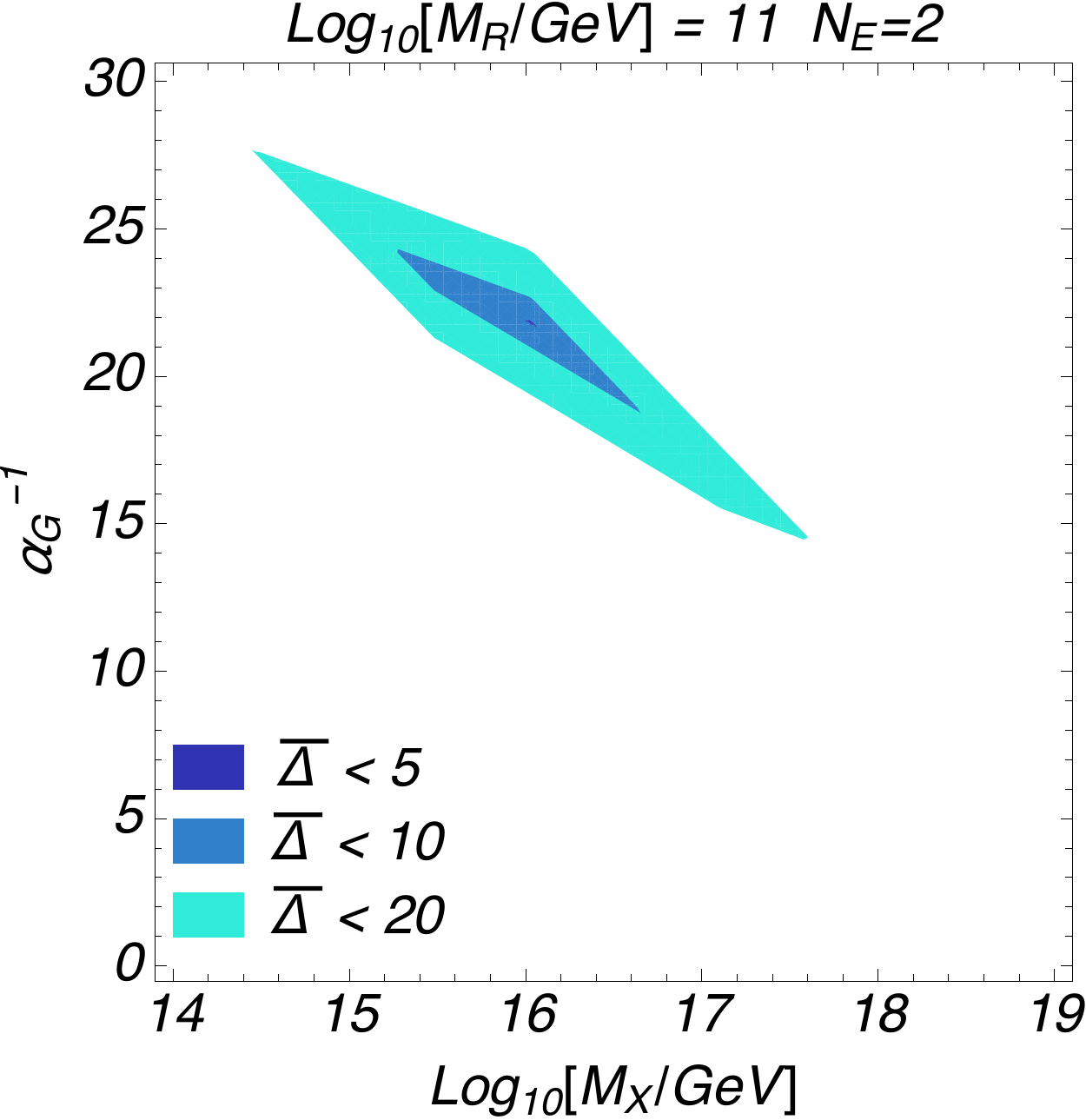}
  \end{minipage}
    \begin{minipage}{.24\linewidth}
 \includegraphics[width=\linewidth]{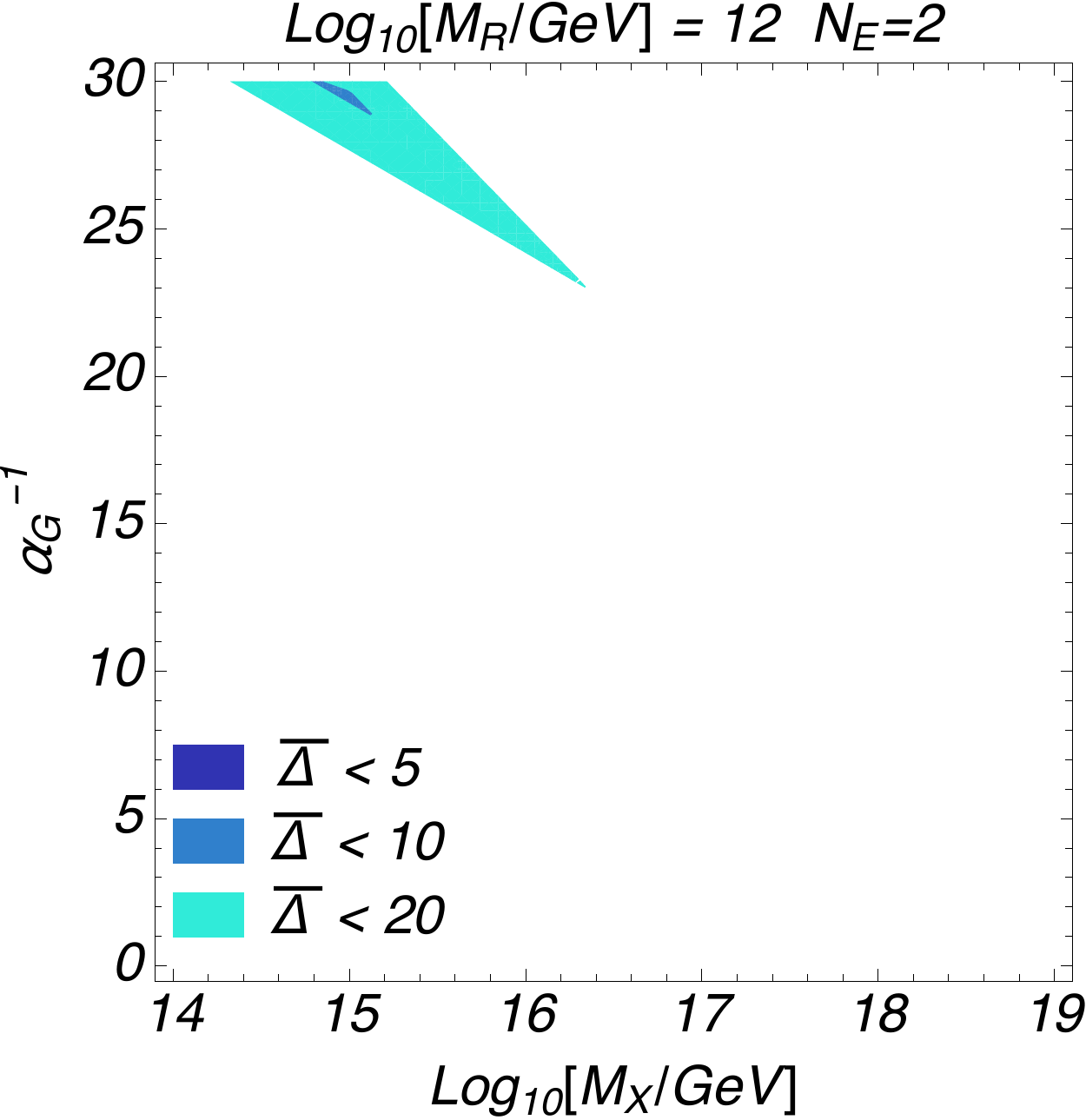}
   \end{minipage}
\caption{\sl \small The quality of the unification $\bar\Delta$ as a function of $(M_X, \alpha_G^{-1})$.
The upper and the lower panels are for $N_E=3$ and $N_E =2$, respectively.
The quality of the unification is reasonably high in the blue shaded region ($\bar\Delta <5$),
while it is moderate in the light-blue shaded region ($\bar\Delta < 10$).
The parameter $\bar{\Delta}$ gets contribution not only from the mass splittings of the GUT multiplets
but also from the mass difference between the GUT particles and $M_X$.
The pink shade region is excluded as $M_X$ is above the Landau pole of $\alpha_{1,2,3}(\mu,M_\R)$.
We confine ourselves to the region 
with $M_{X}\ll M_{\mathrm{Pl}}$,
so that the effective field theory without gravity is valid.
}
\label{fig:unification}
\end{figure}

As a measure of the quality of the unification, we define,
\begin{eqnarray}
\bar{\Delta} \equiv \max_{a=1,2,3}\left[
\Delta_{a}
\right],
\end{eqnarray}
where we take $\mu = \Lambda = M_X$.
The definition of $\bar{\Delta}$ is different from the unification measure $\Delta$ defined in \cite{SO10_LR,SO10_LR2}.
The parameter $\bar{\Delta}$ gets contribution not only from the mass splittings of the GUT multiplets
but also from the mass difference between the GUT particles and $M_X$,
while $\Delta$ in  \cite{SO10_LR,SO10_LR2} purely measures the precision of the unification.

In Fig.~\ref{fig:unification}, we show $\bar{\Delta}$ as a function of $(M_X, \alpha_G^{-1})$ for a given $M_\R$.
The quality of the unification is reasonably high in the blue shaded region ($\bar\Delta <5$),
while it is moderate in the light-blue shaded region ($\bar\Delta < 10$).
The figure shows that a reasonable unification, i.e. $\bar\Delta <5$, is not possible for $M_\R = {\mathcal O}(10^{10})\,$GeV
due to the Landau pole for $N_E=3$.
The figure also shows that the unification is possible for a wide range of the GUT gauge boson mass, 
$M_{X} = 10^{15}$--$10^{17}$\,GeV.
These results should be compared with the previous analyses of the gauge coupling unification  in the LR symmetric model
which preferred $M_\R= {\mathcal O}({10^{10}})$\,GeV and $M_X = {\mathcal O}(10^{17})$\,GeV~\cite{Siringo:2012bc}.
The difference of the results stem from the explicit inclusion of the three flavors of the extra multiplet into the analysis of the RG flow.

For comparison, we also show $\bar{\Delta}$ for $N_E=2$. In this case, the Landau pole is at the very high energy scale and 
does not exclude the parameter region significantly.
For $N_E=2$, more precise unification is achieved for a lower $M_\R$ and 
a higher $M_X$ than the case of $N_E =3$.
In such a parameter region, however, there is a tension with the possibility to obtain the first generation Yukawa couplings 
as the higher dimensional operators suppressed by $M_{\text{GUT}}$.

\section{Proton Lifetime}
\label{sec:Proton}

In the present model, the exchanges of the massive gauge boson in the $(3,2,2)_{-2/3}$ representation, i.e. the $X$-type gauge bosons, induce the proton decay.
Incidentally, the each of the $SU(2)_\R$ doublet component of the $X$ gauge boson
belongs to the adjoint representation 24 and the anti-symmetric representation 10 of the minimal $SU(5)$ GUT
gauge symmetry, respectively.
In general setup of the $SO(10)$ GUT, they have different masses (see e.g. \cite{Haba:2019wwt}), 
while they are common in the LR symmetric model.
The massive gauge boson in the $(3,1,1)_{4/3}$ representation, on the other hand, does not lead 
to the proton decay.

\begin{table}[t]
\bgroup
\def\arraystretch{1.3}
\begin{center}
\begin{tabular}{c|c|c}
$G_\tx{LR}$ &
$G_\tx{SM}$ &
$SU(3)_\C \times U(1)_{em}$  
\\ \hline \hline
\multirow{2}{*}{$\mathcal{O}^{(1)}\equiv(\overline{{L_\L}^c}Q_\L)(\overline{{Q_\R}^c}Q_\R)$} &
\multirow{2}{*}{$2(\overline{{l_\L}^c}q_\L)(\overline{{u_\R}^c}d_\R)$} &
$\mathcal O^1\equiv2(\overline{{e_\L}^c}u_\L)(\overline{{u_\R}^c}d_\R)$   \\ 
& & $\mathcal O^2\equiv2(\overline{{\nu_\L}^c}d_\L)(\overline{{u_\R}^c}d_\R)$\\
\hline
$\mathcal{O}^{(2)}\equiv(\overline{{L_\R}^c}Q_\R)(\overline{{Q_\L}^c}Q_\L)$ &
$(\overline{{e_\R}^c}u_\R)(\overline{{q_\L}^c}q_\L)$ &
$\mathcal O^3\equiv 2(\overline{{e_\R}^c}u_\R)(\overline{{u_\L}^c}d_\L)$  \\ \hline
\end{tabular}
\caption{\sl \small The $B$ and $L$ violating operators mediated by the $X$ gauge boson.}
\end{center}
\egroup
\label{tab:BLbreaking}
\end{table}%

After integrating out the $X$ gauge boson, the gauge interaction of the matter field $F_{16}$ results 
in the $B$ and $L$ breaking operators $\mathcal{O}^{(1,2)}$ in  Table~\ref{tab:BLbreaking}. 
Those operators are reduced to 
\begin{align}
\mathcal{L}_\tx{eff}=\frac{g_G^2}{M_X^2} \left\{
\paren{\overline{{e_\R}^c}u_\R}\paren{\overline{{q_\L}^c}q_\L}+
\paren{\overline{{l_\L}^c}q_\L}\paren{\overline{{u_\R}^c}d_\R}
\right\}+
\frac{g_G^2}{M_{X}^2}
\paren{\overline{{l_\L}^c}q_\L}\paren{\overline{{u_\R}^c}d_\R},
\label{eq:pdecay}
\end{align}
in terms of the $G_\text{SM}$ fields~\cite{Bertolini:2013vta}; see also \cite{Langacker:1980js}.
Below the electroweak symmetry breaking scale, we may decompose it into the proton decay operators in terms of the $SU(3)_\C\times U(1)_\tx{em}$ fields such that $\mathcal{L}_\tx{eff}=C^I \mathcal{O}^I$ 
as in  Table~\ref{tab:BLbreaking}.
In Eq.\,\eqref{eq:pdecay}, we do not take account of the effects of the quark mixing angles~\cite{Ellis:1980jm}.%
\footnote{In the present model, an SM fermion is a linear combination of the spinor $F_{16}$ 
and the extra particles $E_{10}$ and $E_{45}$~\cite{SO10_LR,SO10_LR2}, and therefore we should consider the proton decay operators 
which come from the gauge interactions of the extra particles too, strictly speaking.
However, we have introduced the extra particles to realize the large Yukawa couplings, 
while the Yukawa couplings of the first two generations are small.
Therefore, we expect that the contributions from extra particles are small for the first generation, 
and thus we do not consider contribution from extra particles in this paper.
The proton decay operators which come from the gauge interaction of the extra particle $E_{10}$ are summarized in Ref.~\cite{extra_PD}.}

The partial decay widths for the $p \to \pi^0 e^+$ is given by
\begin{equation}
\Gamma(p \to \pi^0 e^+) \simeq \frac{m_p}{32 \pi} \left\{ 1 - \left( \frac{m_{\pi^0}}{m_p} \right)^2 \right\}^2 \sum_{I=1,3}\left| C^I(m_p) W_0^I \right|^2,
\end{equation}
where $m_p$ and $m_{\pi^0}$ are the proton and the neutral pion masses, respectively, and $W_0^I$ are the proton form factor.
We may safely approximate as $\sum_{I=1,3}\left| C^I(m_p) W_0^I \right|^2=\sum_{I=1,3}\left| C^I(m_p) \right|^2W_0^2$.
In this calculation, $W_0$ for $p \to \pi^0 e^+$ decay mode is $-0.131$\,GeV$^2$ which have been obtained by
lattice simulation~\cite{Aoki:2017puj}.

To calculate the coefficients of the proton decay operators at the proton mass scale $m_p$, 
we have to consider the renormalization factor $A$.
In this paper, we consider the one-loop level renormalization factor from gauge interactions.
Here, we divide the energy region into two parts.
The first region is between the GeV scale and the LR-breaking scale $M_\R$, where
the renormalization factor is written as $A_\tx{long}$.
The second region is between the LR-breaking scale $M_\R$ and the GUT scale $M_{\text{GUT}}$,
where renormalization factor is written as $A_\tx{short}$.
The total renormalization factor $A$ is given by the product of these factors,  $A = A_\tx{long} \times A_\tx{short}$.
We calculate this renormalization factor for each of the proton decay operators $\mathcal{O}^{(1)}$ and $\mathcal{O}^{(2)}$.

The one-loop level renormalization factor for each gauge group is given by
\begin{equation}
A_a=\left( \frac{\alpha_a(M_{\text{start}})}{\alpha_a(M_{\text{end}})}\right)^{-\frac{C_a}{a_a}},\label{renormalization factor A}
\end{equation}
where $M_{\text{end}}>M_{\text{start}}$;
$a_a$ is the coefficient for $\beta$ function for each gauge coupling which are shown in Eqs.~(\ref{eq:beta_coe1}) and  (\ref{eq:beta_coe2}). 
$C_a$ is the factor appearing in the anomalous dimension $\gamma_a$ of the $a$-th gauge interaction for an each proton decay operator:
\begin{equation}
\gamma_a = -2 C_a \frac{g_a^2}{(4 \pi)^2}.
\end{equation}
The coefficient $C_a$ is summarized in Ref.~\cite{Caswell:1982fx}, with which the renormalization factors are given by%
\footnote{We use the six flavor RG equations even below the electroweak scale.
If we instead 
use the three flavor RG below the 
electroweak scale, the $A_{\mathrm{long}}^{(1,2)}$
are slightly enhanced by about $10$\%.
}
\begin{align}
A_\tx{long}^{(1)} &= 
\left( \frac{\alpha_3(1\,\text{GeV})}{\alpha_3(M_\R)} \right)^{-2/a_3}
\left( \frac{\alpha_2(M_Z)}{\alpha_2(M_\R)} \right)^{-\frac{9}{4}/a_2}
\left( \frac{\alpha_1(M_Z)}{\alpha_1(M_\R)} \right)^{-\frac{11}{12}/a_1},\\
A_\tx{long}^{(2)} &= 
\left( \frac{\alpha_3(1\,\text{GeV})}{\alpha_3(M_\R)} \right)^{-2/a_3}
\left( \frac{\alpha_2(M_Z)}{\alpha_2(M_\R)} \right)^{-\frac{9}{4}/a_2}
\left( \frac{\alpha_1(M_Z)}{\alpha_1(M_\R)} \right)^{-\frac{23}{12}/a_1},\\
A_\tx{short}^{(1)} = A_\tx{short}^{(2)} &= 
\left( \frac{\alpha_3(M_\R)}{\alpha_3(M_{\text{GUT}})} \right)^{-2/a_3}
\left( \frac{\alpha_2(M_\R)}{\alpha_2(M_{\text{GUT}})} \right)^{-2\cdot\frac{9}{4}/a_2}
\left( \frac{\alpha_1(M_\R)}{\alpha_1(M_{\text{GUT}})} \right)^{-\frac{1}{4}/a_1}.
\label{eq:RF}
\end{align}
In Eq.\,(\ref{eq:RF}), we double $SU(2)_\L$ contribution to include the contribution from the $SU(2)_\R$ gauge interaction.
For $M_R \simeq 10^{11}$\,GeV, $M_X\simeq 10^{16.5}$\,GeV and $N_E = 3$, for example, we find that the renormalization 
factors are given by,%
\footnote{For this choice, we find $\alpha^{-1}_{1}(M_X, M_\R) \simeq 6.9$, $\alpha^{-1}_{2}(M_X, M_\R) \simeq 7.3$,
and $\alpha^{-1}_{3}(M_X, M_\R) \simeq 6.0$, respectively.} 
\begin{align}
A^{(1)} = A_{\text{long}}^{(1)} A_{\text{short}}^{(1)} \simeq 5.9, \quad
A^{(2)} = A_\tx{long}^{(2)} A_\tx{short}^{(2)} \simeq 6.0.
\end{align}

\begin{figure}[t]
\centering
  \begin{minipage}{.24\linewidth}
 \includegraphics[width=\linewidth]{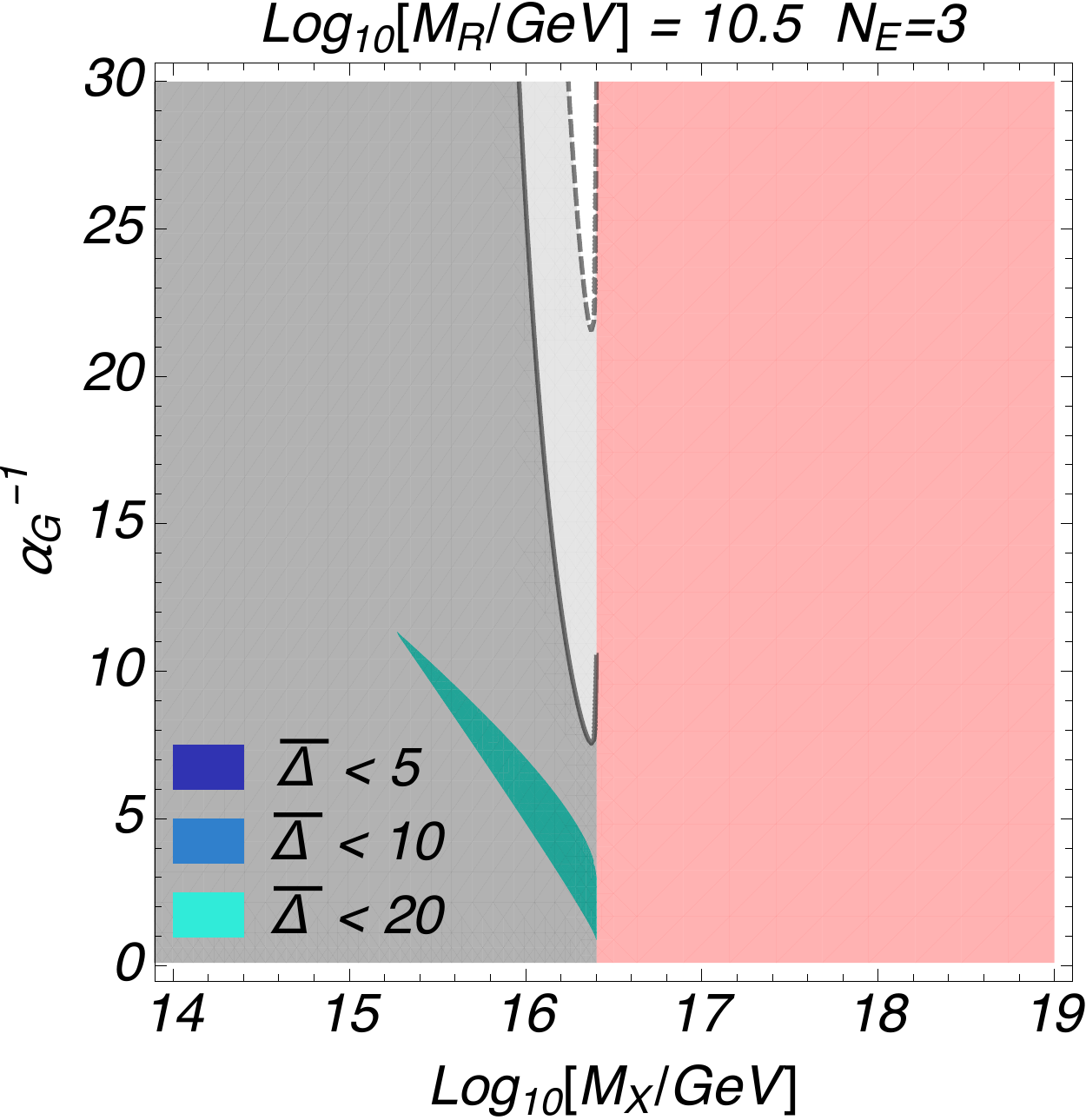}
  \end{minipage}
    \begin{minipage}{.24\linewidth}
 \includegraphics[width=\linewidth]{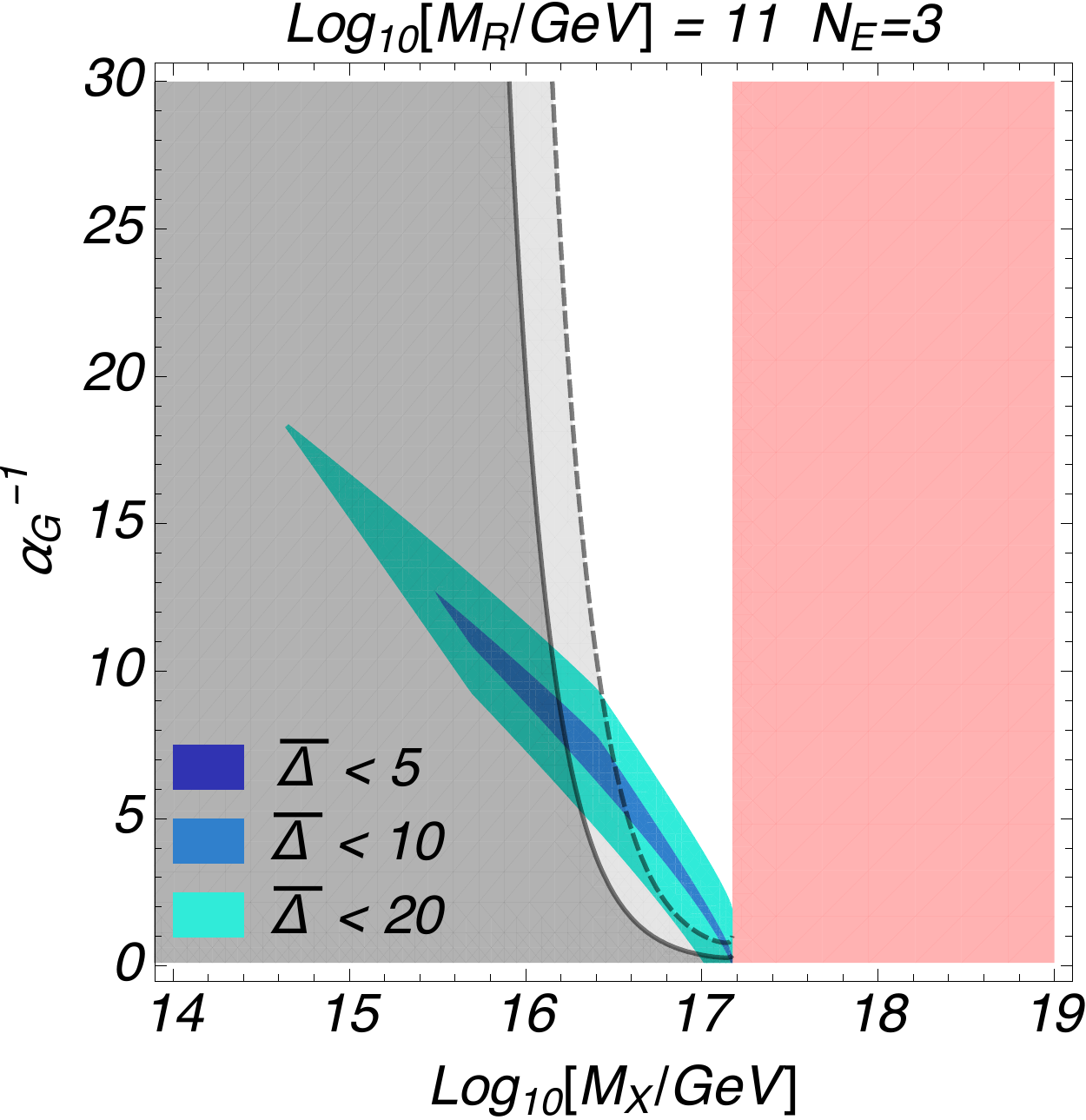}
  \end{minipage}
    \begin{minipage}{.24\linewidth}
 \includegraphics[width=\linewidth]{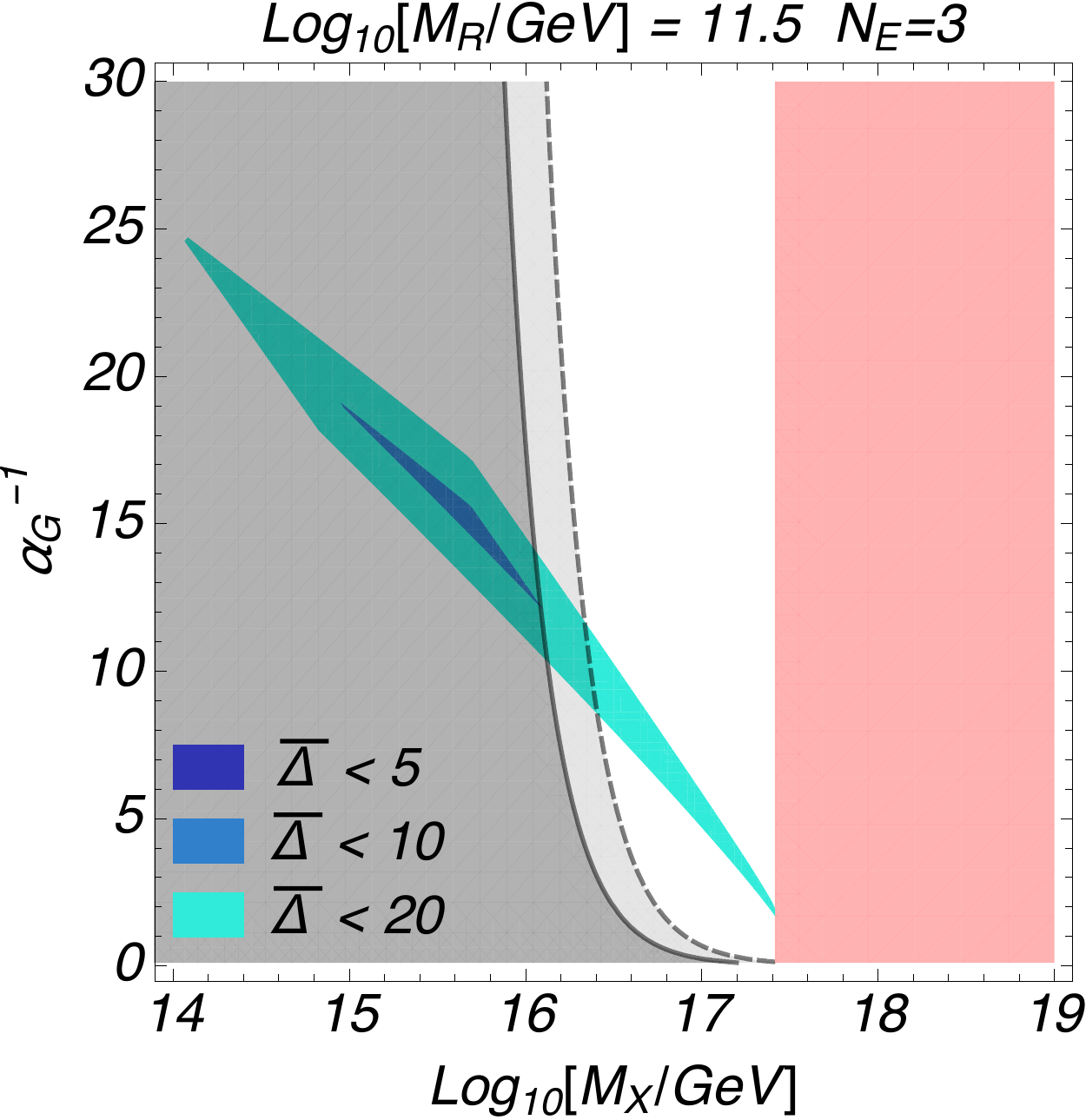}
  \end{minipage}
    \begin{minipage}{.24\linewidth}
 \includegraphics[width=\linewidth]{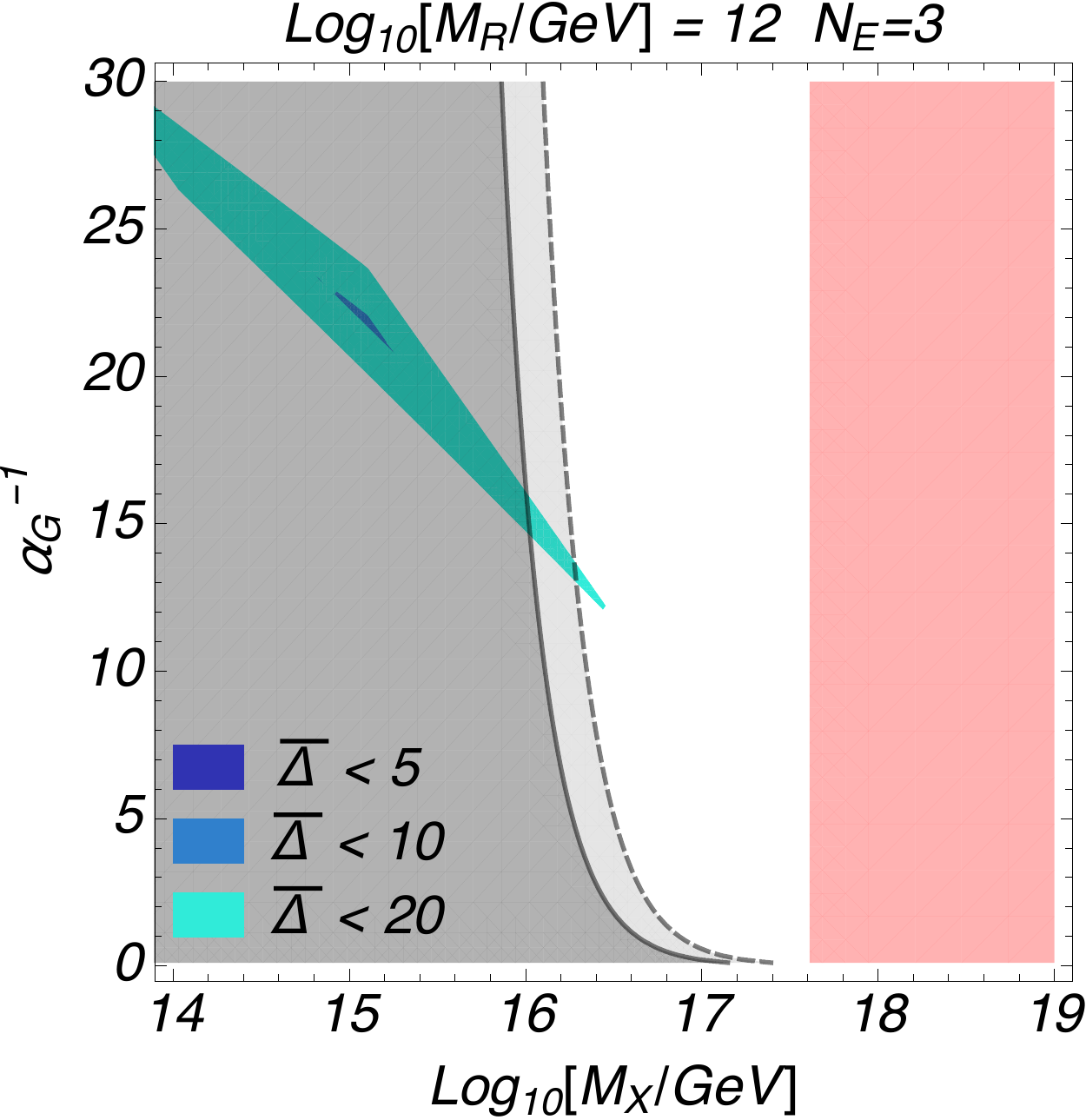}
  \end{minipage}
  \begin{minipage}{.24\linewidth}
 \includegraphics[width=\linewidth]{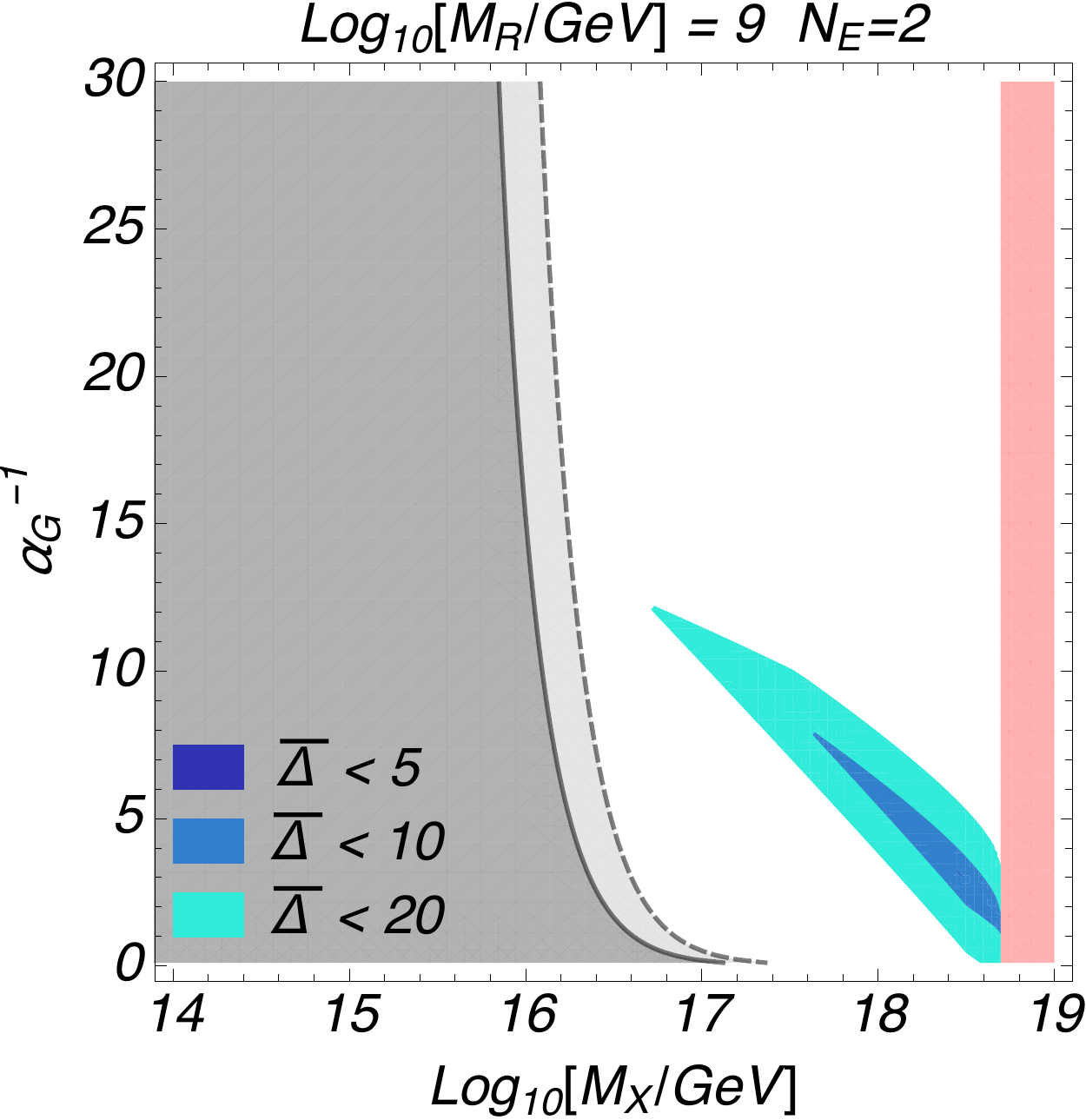}
  \end{minipage}
    \begin{minipage}{.24\linewidth}
 \includegraphics[width=\linewidth]{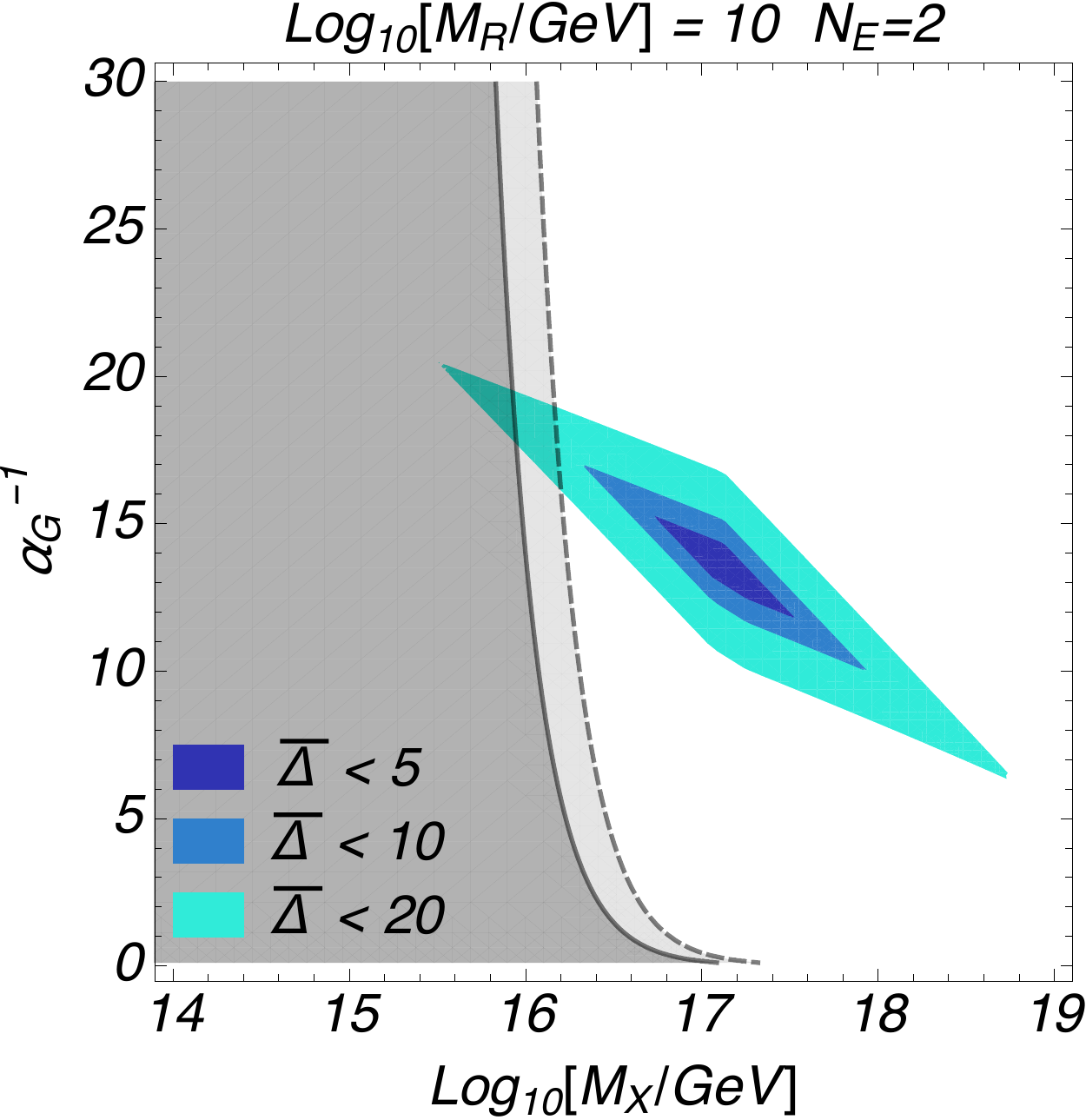}
  \end{minipage}
    \begin{minipage}{.24\linewidth}
 \includegraphics[width=\linewidth]{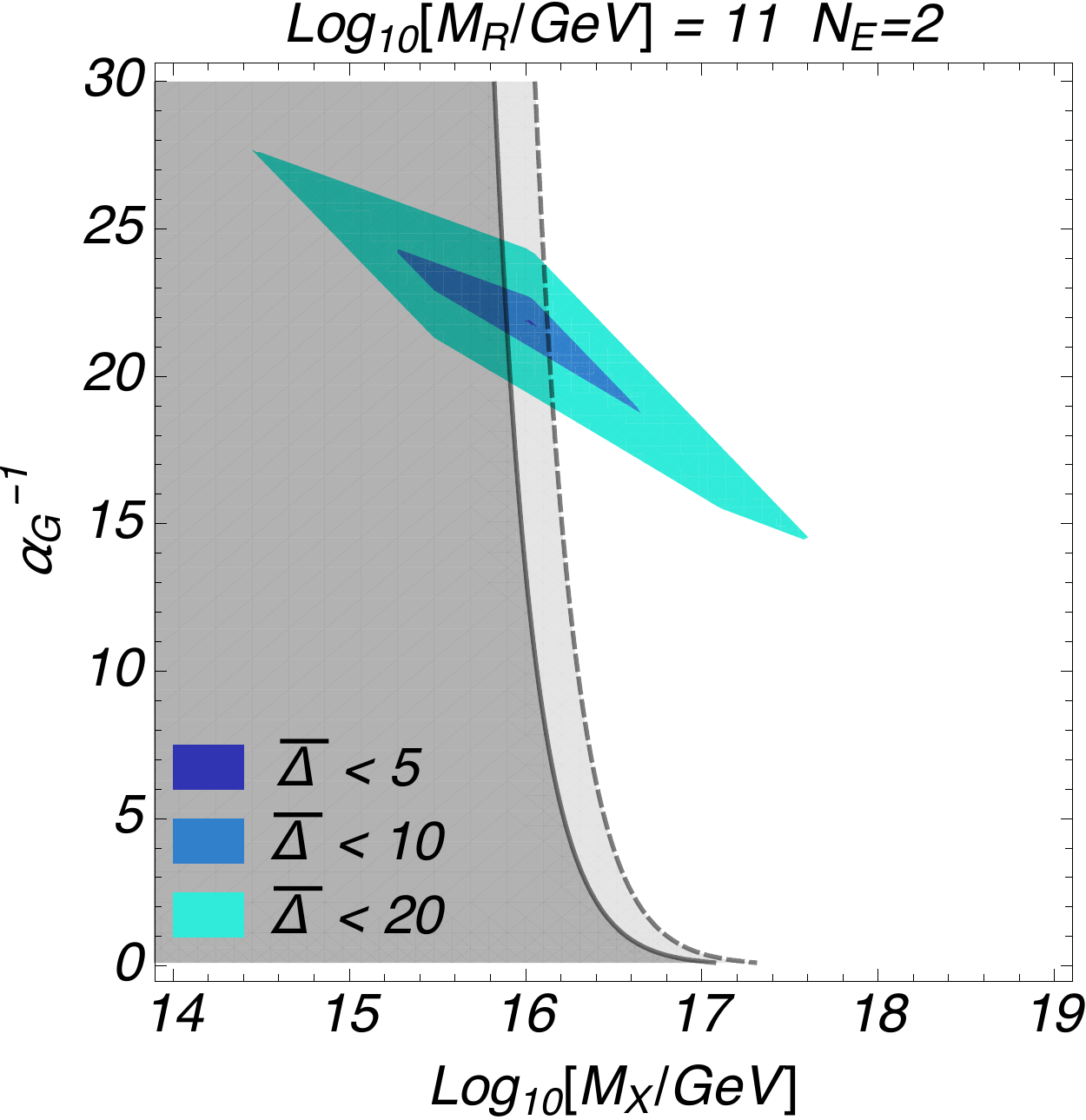}
  \end{minipage}
    \begin{minipage}{.24\linewidth}
 \includegraphics[width=\linewidth]{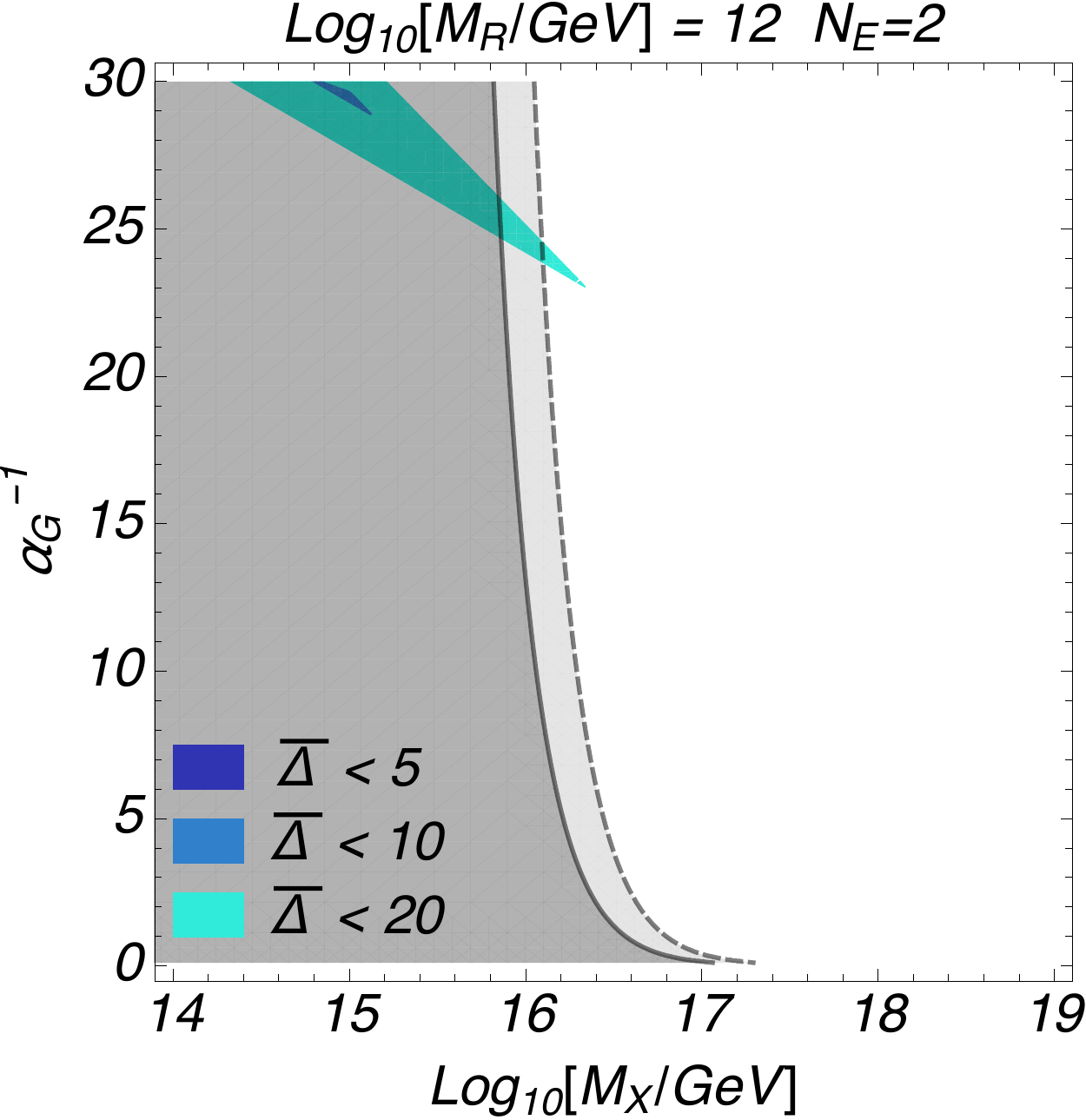}
  \end{minipage}
\caption{\sl \small The black solid and black dashed lines are proton decay constraints
on the  $p \to \pi^0 e^+$ decay mode from current SK limit and the future HK prospect.
The grey shaded region is excluded by the current SK limit 
and the region between black solid line and black dashed line will be explored by the HK experiment.}
\label{fig:proton}
\end{figure}

In Figure~\ref{fig:proton}, we overlay the current limit and the future prospects on the proton lifetime 
for $p \to \pi^0 e^+$ decay mode on Figure~\ref{fig:unification}.
The current limit is the 90\%CL exclusion limit by Super-kamiokande (SK) experiment,  
$1.6 \times 10^{34}$\,years \cite{Miura:2016krn},
which is shown as the black solid line.
The future prospects is the expected exclusion limit at $90$\%C.L. of the Hyper-K (HK) experiment, $1.3 \times 10^{35}$\,years \cite{Abe:2011ts}, which is shown as the black dashed line.
The figure shows that some of the parameter region with moderate coupling unification have 
been excluded by the current SK limit for $M_R \gtrsim 10^{11.5}$\,GeV ($N_E = 3$).
The figure also shows that the HK experiment has a sensitivity to test large portion 
of the parameter space with moderate coupling unification for $M_\R = O(10^{11})$\,GeV for 
$N_E = 2,3$.

\section{Model with Peccei-Quinn Symmetry}
\label{sec:axion}
In the minimal setup with $N_E =3$, we assume that all the SM Yukawa interactions are generated by 
 integrating out the extra vector-like multiplets with masses around the LR-breaking scale.
In this section, we briefly discuss a possibility to generate those masses by the PQ symmetry breaking.
The PQ mechanism is one of the most successful solution to the Strong CP problem~\cite{Peccei:1977hh,Peccei:1977ur}.%
\footnote{Alternatively, the strong CP problem can be solved in the LR symmetric model by imposing space-time parity appropriately; see \cite{SO10_LR,SO10_LR2} and references therein.
}
There, the effective $\theta$-angle of QCD is canceled by the VEV of the pseudo-Nambu-Goldstone boson, 
axion~$a$, which is associated with the spontaneous breaking of the PQ symmetry~\cite{Weinberg:1977ma,Wilczek:1977pj}.
The axion model not only solve the strong CP problem, but also provides 
a good candidate for cold dark matter~\cite{Dine:1982ah,Abbott:1982af,Preskill:1982cy,Preskill:1991kd}; 
see also Ref.~\cite{Khlopov}.
In fact, the axion dark matter model is successful when the PQ breaking scale is of $10^{11\text{--}12}$\,GeV,
which is close to the LR-breaking scale discussed in this paper; see \cite{Kawasaki:2013ae} for review.
This coincidence motivates us to see how it is successful to the mass scale of the extra vector-like fermions 
with the PQ breaking scale.

For this purpose, let us introduce a gauge singlet complex scalar field, $P$, which breaks the PQ symmetry at an 
intermediate scale. The PQ charge of $P$ is defined to be $1$.
Below the PQ breaking scale, the axion appears as a phase component of $P$,
\begin{eqnarray}
P = \frac{1
}{\sqrt{2}} f_{a}
e^{ia/f_{a}},
\end{eqnarray}
where $f_a$ is the decay constant of the axion.
The PQ symmetry is realized by the shift of $a$,
\begin{align}
{a\ov f_a}\to {a'\ov f_a} = {a\ov f_a} + \alpha,  \quad (\alpha \in {\mathbb R})
\end{align}
where the domain of the axion is given by $a/f_a = [-\pi,\pi)$.

To generate the extra fermion masses at the PQ scale, we assume that $P$ couples to $E_{10,45}$ via,
\begin{eqnarray}
{\cal L} = k_{10} P E_{10} E_{10} + k_{45}  P E_{45} E_{45}+\tx{h.c,}
\end{eqnarray}
where $k_{10,45}$ are the coupling constants.
Here, we assume that the PQ charges of $E_{10}$ and $E_{45}$ are $-1/2$.
In this case, the interaction terms in Eq.\,\eqref{Eq:extra_cont} impose that 
the PQ charges of $F_{16}$ are $1/2$, while that of $H_{16}$ is vanishing.%
\footnote{We may consider a model in which $E_{10}$ and $E_{45}$ have the opposite 
PQ charges. In this case the PQ charge of $F_{16}$ is vanishing,
although the domain wall number is again $N_{\text{DW}} = -21$.
 }

With these charge assignments, we find that the anomalous axion coupling to QCD is given by,
\begin{align}
\label{eq:QCD}
&  \mathcal{L}= \frac{g_3^2}{32\pi^2 } N_{\text{DW}}\frac{a}{f_a} G\tilde{G},
&  N_{\text {DW}}= (2N_{F_{16}}-N_{E_{10}} - 8N_{E_{45}} ) = -21.
\end{align}
Here, $G$ and $\tilde G$ are the QCD field strength and its hodge dual, respectively.
$N_{F_{16}}=3$ is the number of generation of the SM fermions,
and $N_{E_{10}}=N_{E_{45}} = N_E = 3$.
The Lorentz and color indices are suppressed. 
Below the QCD scale the anomalous coupling of the axion to QCD in Eq.\,\eqref{eq:QCD} leads 
to a non-vanishing axion potential and the axion settles down to its minimum which solves 
the strong CP problem.%
\footnote{Here, the origin of the axion field space is taken to be the one at which the effective $\theta$-angle of QCD is vanishing
without loss of generality.}
As both the extra fermions as well as the SM fermions possesses the PQ charges, this model is in between the 
KSVZ~\cite{Kim:1979if,Shifman:1979if} and DFSZ~\cite{Zhitnitsky:1980tq,Dine:1981rt} invisible axion models, and is in principle distinguishable from these models.

The coherent oscillation of the axion turns into the dark matter density~\cite{Turner:1985si},
\begin{align}
\Omega_a h^2 \simeq 0.18 \left(\frac{{\mit \Delta}a_i}{F_{\text{eff} }} \right)^2
\left( \frac{F_{\text{eff}}}{10^{12}\,\text{GeV}}\right)^{1.19},
\end{align}
where we have defined $F_{\text{eff}} = f_a/N_{\text{DW}}$.
${\mit \Delta}{a_i}/{F_{\text{eff}}} \in [-\pi,\pi)$ denotes the initial misalignment angle of the axion 
from the $N_{\text{DW}}$ degenerate CP conserving vacua.
Therefore, the axion dark matter scenario is successful for $F_{\text{eff}}\sim 10^{11\tx{--}12}$\,GeV
for a typical initial misalignment angle.
In this present model, 
the PQ breaking scale is given by $f_a = N_\tx{DW} F_{\text{eff}}$, the axion dark matter prefers 
the PQ breaking scale at $f_a \sim 10^{12\tx{--}13}$\,GeV.
Accordingly, we find that the extra multiplet masses at $M_R \sim 10^{11}$\,GeV can be provided 
for $k_{10,45} \sim 10^{-(1\text{--}2)}$
consistently with the axion dark matter scenario.
It should be emphasized that this scenario does not work for $N_E= 2$ since
the higher dimensional operator to generate the SM Yukawa interactions of the first generation
explicitly break the PQ symmetry.%
\footnote{We may consider the PQ symmetry which is spontaneously broken at the cutoff scale
even for $N_E = 2$.
In such a case, however, the axion dark matter scenario is not successful.}

We argue that the axion in our setup is within the reach of future detection.
Due to the non-vanishing axion potential, the axion get a mass given by~\cite{Gorghetto:2018ocs}\
\begin{eqnarray}
m_a \simeq
5.7\, \mu {\rm eV} \left (\frac{10^{12}\,{\rm GeV}}{F_{\rm eff}} \right).
\end{eqnarray}
The axion also couples to photons through the electromagnetic anomaly $N_{\mathrm{QED}}$ and thorough the mixing with neutral mesons.
Many on-going and future axion search experiments utilize the axion-photon coupling, which is parameterized as 
\begin{eqnarray}
\mathcal{L} \supset \frac{g_{a \gamma \gamma}}{4} a F \tilde F ,
\end{eqnarray}
with~\cite{diCortona:2015ldu}
\footnote{The ratio $N_{\mathrm{QED}}/N_{\mathrm{DW}}=8/3$ 
is a generic feature of the GUT consistent PQ charge assignment.}
\begin{eqnarray}
g_{a\gamma\gamma} = \frac{\alpha_{\rm EM}}{2\pi F_{\rm eff}} \left( \frac{N_{\mathrm{QED}}}{N_{\mathrm{DW}}} - 1.92 (4) \right)
 = \frac{\alpha_{\rm EM}}{2\pi F_{\rm eff}} \left( \frac{8}{3} - 1.92 (4) \right).
\end{eqnarray}
Note that $g_{a\gamma\gamma}$ in our model is equivalent to that in the DFSZ axion model~\cite{Zhitnitsky:1980tq,Dine:1981rt}, which is already excluded by the current ADMX experiment for $m_a \simeq 2.7$\,-\,$3.3$\,$\mu$eV~\cite{Du:2018uak,Braine:2019fqb}. 
The higher mass range of $m_a$ up to 400\,$\mu$eV (corresponding to $F_{\rm eff} \sim 10^{11}$\,GeV) is expected to be covered by future cavity haloscopes such as ADMX~\cite{Du:2018uak}, CULTASK~\cite{Petrakou:2017epq} and MADMAX~\cite{TheMADMAXWorkingGroup:2016hpc}; see also Ref.~\cite{Graham:2015ouw,Lawson:2019brd}.

Several comments are in order.
The axion potential induced by the anomalous QCD coupling in Eq.\,\eqref{eq:QCD} possesses
${\mathbb Z}_{N_{\text{DW}}}$ discrete symmetry in the domain of the axion $a/f_a \in [-\pi,\pi)$, or equivalently in
$a/F_{\text{eff}} \in N_{\text{DW}} \times [-\pi,\pi)$.
The discrete symmetry is spontaneously broken by the VEV of the axion.
Thus, the domain wall formation takes place after the onset of the coherent oscillation of the axion,
if the initial misalignment angle in each Hubble volume of the Universe at that time is random.
Once the domain walls are formed, they immediately dominate the Universe, which conflicts with the Standard Cosmology.
To avoid this problem, we need to assume that the PQ symmetry breaking takes place before inflation
and never gets restored after inflation.
Under this assumption, the initial misalignment angle of the axion is uniform in the entire Universe,
and hence the axion sits in 
the same sub-domain and evades the formation of the domain wall.

We mention that the large domain wall number, $N_{\mathrm{DW}} = -21$, is advantageous to avoid 
the PQ-symmetry restoration, since the actual PQ breaking scale is an order of magnitude larger than the effective decay constant $F_{\text{eff}}$ appropriate for the axion dark matter scenario, i.e. $F_{\text{eff}} \sim 10^{11\tx{--}12}$\,GeV. 
Therefore, the present model can be consistent with
a cosmological scenario with higher 
reheating temperature than in the conventional axion
dark matter models.
In this sense, the present model can be more easily consistent with the thermal leptogenesis scenario~\cite{Fukugita:1986hr} which requires a rather high reheating 
temperature, $T_R \gtrsim 10^{9\mbox{-}10}$\,GeV~\cite{Giudice:2003jh,Buchmuller:2005eh,Davidson:2008bu}.%
\footnote{For a given reheating temperature $T_R$, the maximal 
temperature of the Universe of the thermal plasma during the
inflaton dominated era is in general much higher than $T_R$
up to $T_{\rm Max}\sim (T_R^2 H_I M_{\text{Pl}})^{1/4}$~\cite{Kolb:1990vq,Yokoyama:2004pf,Harigaya:2013vwa}.
Here, $H_I$ is the Hubble parameter during inflation and $M_{\text{P}}$ is the reduced Planck scale.}

As another comment, the massless axion fluctuates 
quantum mechanically during inflation,
which leads to the isocurvature fluctuation of the axion dark matter density
when the PQ symmetry breaking takes place before inflation.
The dark matter isocurvature fluctuation have been 
severely constrained by the precise measurements of the 
cosmic microwave background~\cite{Akrami:2018odb}. 
The amplitude of the isocurvature fluctuation is proportional to 
the Hubble parameter during inflation, $H_I$.
As a result, $H_I$ is constrained from above as $H_I \lesssim 10^{7\mbox{--}8}$\,GeV 
to avoid the current constraint; see e.g. Ref.~\cite{Kawasaki:2013ae,Kawasaki:2018qwp}.
Therefore, the present scenario with the axion dark matter can be refuted if the primordial $B$-mode polarization in the cosmic microwave background 
is discovered in near future; see, 
e.g. Ref.~\cite{Abazajian:2016yjj,Lavrelashvili:1987jg}.

Finally, let us comment on the origin of the 
PQ symmetry.
By definition, the $U(1)$ PQ symmetry cannot be
an exact symmetry as it is explicitly broken 
by the QCD anomaly.
Besides, it is also argued that any global symmetries 
are broken by quantum gravity effects~\cite{Hawking:1987mz,Lavrelashvili:1987jg,Giddings:1988cx,Coleman:1988tj,Gilbert:1989nq,Banks:2010zn}. 
When explicit breaking terms exist,
the effective $\theta$ angle of QCD 
is non-vanishing even in the presence of the axion,
which spoils the PQ mechanism.
For example, if the PQ symmetry is completely broken 
by the quantum gravity effects, it is expected that there 
should be a PQ breaking term at least,
\begin{align}
{\cal L}_{\text{PQ--breaking}} = \frac{P^5}{M_{\text{Pl}}} + \tx{h.c.}
\end{align}
which drastically affects the axion potential and spoils the PQ mechanism.

In the present model, however, we may regard 
that the discrete $\mathbb{Z}_{2N_{\rm DW}}$ symmetry 
to be a discrete gauge symmetry as it can satisfy
the anomaly free conditions~\cite{Csaki:1997aw}.%
\footnote{Here, we normalize the charges of the discrete symmetry so that the extra vector-like multiplets 
have a charge $-1$. Accordingly, the PQ breaking field $P$ possesses the discrete charge $2$.}
If ${\mathbb{Z}}_{2N_{\rm DW}}$ symmetry is a gauge symmetry, the lowest dimensional operator which breaks 
the $U(1)$ PQ symmetry but is invariant under 
the $\mathbb{Z}_{2N_{\rm DW}}$ gauge symmetry is given by,
\begin{align}
    {\cal L}_{\text{PQ--breaking}} = \frac{P^{21}}{M_{\text{Pl}}^{17}} + \tx{h.c.}
\end{align}
which is highly suppressed and does not spoil the PQ mechanism; see e.g.\ Ref.~\cite{Carpenter:2009zs}. 
This argument strengthens the PQ mechanism in the 
present model.%
\footnote{We discus the domain wall problem in Appendix~\ref{sec:discrete}.
As another possible justification of the 
$U(1)$ PQ symmetry, we may consider a $U(1)$ gauge symmetry with an accidental global $U(1)$ PQ symmetry~\cite{Barr:1992qq}, where $E_{10}$ and $E_{45}$ couple 
to a different PQ charged complex scalars: see Refs.~\cite{Fukuda:2017ylt,Fukuda:2018oco,Ibe:2018hir} for details.}

\section{Summary}
In this paper, we have investigated the proton lifetime in the $SO(10)$ GUT which is broken down by the VEV of $H_{45}$ to 
the minimal LR-symmetric gauge group $SU(3)_\C\times SU(2)_\L\times SU(2)_\R\times U(1)_Y$, which is in turn broken at the intermediate LR-breaking scale $M_\R$ by the $SU(2)_\R$ doublet Higgs that is a part of $H_{16}$.
The $SU(2)_\L$ doublet component 
of the same $H_{16}$ field eventually plays the
role of the SM Higgs doublet.
Due to the absence of the bi-doublet Higgs boson, the 
LR-breaking scale is determined to
be at around $10^{10\text{--}12}$\,GeV
in order to achieve the gauge coupling unification.

As a notable feature of the model, it requires extra vector-like fermions to generate 
the SM Yukawa interactions.
Such extra multiplets 
affect the RG flow, and lower the unification scale down to $M_X\lesssim 10^{17}$\,GeV from that expected in Refs.~\cite{Chang:1984qr,Siringo:2012bc} by a factor a few or so.
We have also found that the Wilson coefficients of the proton decay operators are considerably larger 
than those in the minimal $SU(5)$ GUT model.
As a result, the proton decay rate is enhanced and we find that some portion of the parameter space 
consistent with the gauge coupling unification can be tested by the Hyper-K experiment 
thorough the proton decay search even when the GUT gauge boson mass is in the range $10^{16\tx{--}17}$\,GeV.

We also discussed a possibility to generate the mass of the extra vector-like multiplets by the PQ symmetry breaking.
We found that the axion dark matter scenario and the present model can be successfully combined for the model with 
$N_E = 3$.
This combination can be tested by the proton decay search, the axion search and the search for the primordial $B$-mode fluctuation in the cosmic microwave background.

\section*{Acknowledgments}
\noindent We thank Shigeki Matsumoto, Kyohei Mukaida, and Kenichi Saikawa for useful discussion.
We are grateful to Keisuke Harigaya for valuable comments and cross-checks.
This work is supported in part by JSPS KAKENHI Grant Nos.~15H05889, 16H03991, 17H02878, 18H05542 (M.I.), 19H01899 (K.O.)  15H05889, 15K21733, and 17H02875 (N.Y.); World Premier International Research Center Initiative (WPI Initiative), MEXT, Japan (M.I.); National Natural Science Foundation of China (NNSFC) under Contracts Nos.~11675061, 11775092, 11521064, and 11435003 (Y.M.); and the European Union via the Advanced ERC grant SM-grav, No 669288 (Y.H.).

\appendix

\section*{Appendix}
\section{Discrete Gauge Symmetry and the Domain Wall Problem}
\label{sec:discrete}
In Sec.~\ref{sec:axion}, we considered a discrete gauge symmetry which 
explains the origin of the approximate global $U(1)$ PQ symmetry.
In this appendix, we briefly comment on the domain wall problem 
in the presence of the discrete gauge symmetry behind the PQ symmetry. 
In this set up, the $N_{\mathrm{DW}}$ axion domains in $a/F_{\mathrm{eff}} = N_{\mathrm{DW}}\times [-\pi,\pi)$ are gauge equivalent with each other, and hence,
the axion domain wall configurations which connects different
domains are not completely stable.
As we will see, however, the axion domain wall problem remains 
even in the model with the discrete gauge symmetry.

To make our discussion concrete, 
let us assume that the discrete $\mathbb{Z}_{N_{DW}}$ gauge 
symmetry originates from a $U(1)$ gauge symmetry broken by the VEV of a complex scalar $\Phi$ 
whose gauge charge is large, $N_{\mathrm{DW}}\gg 1$.
Note that this $U(1)$ gauge symmetry is different from the global $U(1)$ PQ symmetry.
The $U(1)$ gauge charge of the PQ breaking field
$P$ is $1$ as in Sec.\,\ref{sec:axion}.%
\footnote{To make the $U(1)$ gauge symmetry anomaly free,
we need to introduce additional SM charged fermions 
(see e.g. \cite{Fukuda:2017ylt,Fukuda:2018oco,Ibe:2018hir}), although
they do not affect the following discussion. Here, we only pay attention to the complex scalars where $\mathbb{Z}_{N_\mathrm{DW}}$
in stead of $\mathbb{Z}_{2N_\mathrm{DW}}$
is good enough for the following discussion.
}
The VEV of the PQ breaking field $P$ eventually breaks the $\mathbb{Z}_{N_{\mathrm{DW}}}$ symmetry.

In this model, the stable topological defect is not the domain wall but the local strings which are associated with the spontaneous $U(1)$ gauge symmetry breaking.
For example, a cosmic local string around which the phase of $\Phi$ winds from $0$--$2\pi$ are expected to be formed when $\Phi$ obtains a VEV at a very high energy scale.
The phase of the PQ breaking field $P$ is changed by $2\pi/N_{\mathrm{DW}}$ under the parallel transport
around this local string, which corresponds to the 
Aharanov-Bohem effect.

Now let us assume that the spontaneous symmetry breaking of the $U(1)$ gauge symmetry by the 
VEV of $\Phi$ takes place well before inflation, while the approximate global PQ symmetry breaking  occurs 
after inflation.
In this case, the cosmic local strings that are formed when $\Phi$ obtains a VEV have been diluted away by cosmic inflation.
After inflation, the cosmic temperature decreases below the PQ breaking scale.
Then, associated with the spontaneous breaking of the approximate global $U(1)$ PQ symmetry, a few cosmic global strings are expected to be formed in each Hubble volume.
Note that these global strings are different from the ones diluted away during the inflation.
When we turn around the global string, the phase of the PQ field $P$
takes values from $0$ to $2\pi$ when the winding number is one,
and hence, the axion field takes values from $0$ to $ f_a \times 2 \pi = N_{\mathrm{DW}}\times F_{\mathrm{eff}}\times 2\pi$.

Around the global string, the $[0,N_{\mathrm{DW}}\times F_{\mathrm{eff}}\times 2\pi)$ region has $N_\text{DW}$ domains that are gauge equivalent under the $\mathbb Z_{N_\tx{DW}}$.
Since the approximate $U(1)$ PQ symmetry is highly protected by the 
$\mathbb{Z}_{N_{\tx{DW}}}$ symmetry, the tension of the domain walls connecting the $N_{\tx{DW}}$ domains is negligibly small.
Therefore, we have no domain wall problem associated with the $\mathbb{Z}_{N_{\tx{DW}}}$ symmetry breaking by $\langle P\rangle\neq 0$.
When the cosmic temperature decreases further, the cosmic global string networks follow the so-called scaling solution where the number of the cosmic global strings in each Hubble volume at that time remains of $\order{1}$; see e.g.~\cite{Vilenkin:2000jqa}.
When the axion potential is generated at around the QCD scale $\Lambda_{\mathrm{QCD}}$,
potential barriers appear around each global string which
result in $N_{\mathrm{DW}}$ domain walls whose boundary is the global string.

As mentioned earlier, each domain wall attached to the global string connects different domains which are gauge equivalent under the 
discrete $\mathbb{Z}_{N_\mathrm{DW}}$ symmetry.
Therefore, this domain wall is not 
completely stable.
In fact, each wall can be punctured by a loop of the earlier mentioned local string, around which the phase of $\Phi$ winds from $0$--$2\pi$, since this local string connects the different axion domains without potential barrier.
Once the domain wall is punctured, the loop of local string expands on the domain wall, and the domain wall disappears eventually.
The rate of such a puncturing process, however, is highly suppressed,
since the formation of the loop of the local string is suppressed by 
$e^{-|\langle{\Phi}\rangle|^4/\Lambda_{\mathrm{QCD}}^2F_{\mathrm{eff}}T}$ at a temperature below the QCD scale: $T\lesssim \Lambda_{\mathrm{QCD}}$.\footnote{
For a loop of a radius $\ell$, the removed wall energy is $\sim\ell^2\times m_a F_\tx{eff}^2$, while the string energy cost is $\sim\ell\times|\langle\Phi\rangle|^2$. Therefore we need at least $\ell\gtrsim|\langle\Phi\rangle|^2/m_a F_\tx{eff}^2$. Multiplying the string tension $|\langle\Phi\rangle|^2$ and putting $m_a\sim \Lambda_\tx{QCD}^2/F_\tx{eff}$, the lowest mass of the string loop which can puncture the domain wall is $|\langle\Phi\rangle|^4/\Lambda_\tx{QCD}^2F_\tx{eff}$.
}
As a result, the domain wall is virtually stable below the QCD scale
and they immediately dominate over the energy density of the universe, which causes the domain wall problem.%
\footnote{For a study of cosmic string formation  when both $\Phi$ and 
$P$ obtain VEVs after inflation, 
see Ref.~\cite{Hiramatsu:2019tua}.}

\end{document}